%% file: BirkCAP.tex
\documentclass[a4paper,12pt,twoside]{article}
\usepackage[british]{babel}
\usepackage{amssymb,amsmath,amsfonts,verbatim,xkeyval,bm}
\usepackage{graphicx}
\usepackage{subfigure}
\usepackage{ wasysym } %per il simbolo pallino nero

\newtheorem{theorem}{Theorem}[section]
\newtheorem{lemma}[theorem]{Lemma}
\newtheorem{proposition}[theorem]{Proposition}

\newenvironment{proof}[1][Proof]{\begin{trivlist}
\item[\hskip \labelsep {\bfseries #1}]}{\end{trivlist}
{\nobreak\quad\nobreak\hfill\nobreak{$\blacksquare$}}}

\def\vet#1{{\boldsymbol #1}}
\def\build#1_#2^#3{\mathrel{
\mathop{\kern 0pt#1}\limits_{#2}^{#3}}}
\def\reali{\mathbb{R}}
\def\complessi{\mathbb{C}}
\def\naturali{\mathbb{N}}
\def\interi{\mathbb{Z}}

\def\toro{\mathbb{T}}

\def\Cscr{\mathcal{C}}
\def\Dscr{\mathcal{D}}
\def\Escr{\mathcal{E}}
\def\Fscr{\mathcal{F}}
\def\Gscr{\mathcal{G}}
\def\Hscr{\mathcal{H}}

\def\Lscr{\mathcal{L}}
\def\Mscr{\mathcal{M}}
\def\Nscr{\mathcal{N}}
\def\Oscr{\mathcal{O}}
\def\Pscr{\mathcal{P}}
\def\Qscr{\mathcal{Q}}
\def\Rscr{\mathcal{R}}
\def\Sscr{\mathcal{S}}

\def\Zscr{\mathcal{Z}}
\def\epsilon{\varepsilon}
\def\rho{\varrho}
\def\phi{\varphi}

\def\poisson#1#2{\left\{ #1,\,#2 \right\}}

\def\lie#1{\Lscr_{#1}}
\def\Lie#1{\Lscr_{#1}}

\def\wideitem#1{\par\hangindent\itemindent
   \noindent\hbox to\parindent{\hfil{#1}\enspace}\ignorespaces}
\def\biwideitem#1{\par\hangindent 3.\itemindent
   \noindent\hbox to 3.\parindent{\hfil{#1}\enspace}\ignorespaces}

\def\mgot{{\mathfrak m}}
\def\Mgot{{\mathfrak M}}

\def\Rgot{{\mathfrak R}}
\def\Sgot{{\mathfrak S}}

\title{\bf Computer-assisted estimates for Birkhoff normal forms\thanks{ {\it 2010 Mathematics Subject
      Classification.}  Primary: 37J40; Secondary: 37N05,
    70--08, 70F07, 70H08. {\it Key words and
      phrases:} computer-assisted proofs, normal form methods, effective stability, Hamiltonian
    systems, Celestial Mechanics.}}

\author{
{\bf CHIARA CARACCIOLO}\\
{\small Dipartimento di Matematica, 
Universit\`a degli Studi di Roma ``Tor Vergata'',}\\
{\small via della Ricerca Scientifica 1, 00133\ ---\ Roma (Italy).}\\
{\bf UGO LOCATELLI}\\
{\small Dipartimento di Matematica, 
Universit\`a degli Studi di Roma ``Tor Vergata'',}\\
{\small via della Ricerca Scientifica 1, 00133\ ---\ Roma (Italy).}\\
{\small e-mails:
  {\tt caraccio@mat.uniroma2.it, locatell@mat.uniroma2.it,}}\\
}

\begin{document}
\maketitle

% *** Le due righe seguenti sono probabilmente da togliere ***
\selectlanguage{british}
\thispagestyle{empty}

\begin{abstract}
  Birkhoff normal forms are commonly used in order to ensure the so
  called ``effective stability'' in the neighborhood of elliptic
  equilibrium points for Hamiltonian systems. From a theoretical point
  of view, this means that the eventual diffusion can be bounded for
  time intervals that are exponentially large with respect to the
  inverse of the distance of the initial conditions from such
  equilibrium points. Here, we focus on an approach that is suitable
  for practical applications: we extend a rather classical scheme of
  estimates for both the Birkhoff normal forms to any finite order and
  their remainders. This is made for providing explicit lower bounds
  of the stability time (that are valid for initial conditions in a
  fixed open ball), by using a fully rigorous computer-assisted
  procedure.  We apply our approach in two simple contexts that are
  widely studied in Celestial Mechanics: the H\'enon-Heiles model and
  the Circular Planar Restricted Three-Body Problem. In the latter
  case, we adapt our scheme of estimates for covering also the case of
  resonant Birkhoff normal forms and, in some concrete models about
  the motion of the Trojan asteroids, we show that it can be more
  advantageous with respect to the usual non-resonant ones.
\end{abstract}

\bigskip

%%%%%%%%%%%%%%%%%%%%%%%%%%%%%
% *** La citazione del Liga e' l'ultima cosa da togliere, perche' porta bene
%\markboth{Non siam tre piccoli Lombardin, non siamo tre porcellin $\ldots$}{
%\markboth{``Leave me alone, I must disassociate from you... I am escaping
%from your grip", Muse}{
%\markboth{``Ooo $1$, $2$, $3$, $4$, fire's in your eyes$\ldots$ And this chaos, it defies imagination$\dots$"}{
 % ``$\ldots$ Ooo $5$, $6$, $7$ minus $9$ lives$\ldots$ You've arrived at Panic Station!", Muse}
\markboth{C. Caracciolo, U. Locatelli}{Computer-assisted estimates for Birkhoff normal forms}
%%%%%%%%%%%%%%%%%%%%%%%%%%%%%

\section{Introduction}\label{sec:intro}
The Birkhoff normal form represents a milestone in the historical
development of the Hamiltonian perturbation theory, because it has
been designed in the simplest possible way so as to (at least) partially
escape the results obtained by Poincar\'e for what concerns both the
non-existence of the first integrals and the ubiquity of chaos
(see~\cite{Poincare-1892}). Indeed, it was introduced about one
century ago (see~\cite{Whittaker-16},
\cite{Cherry-24}--\cite{Cherry-24.1} and~\cite{Birkhoff-27}) in the
framework of the dynamics in the neighborhood of an elliptic
equilibrium point; in section~\ref{sec:birk} such a context is
recalled by a set of proper definitions.  More recently, the scheme of
analytical estimates has been further improved in order to ensure that the
stability time is exponentially big with respect to (some fractional
power of) the inverse of the distance from the equilibrium point
(see~\cite{Littlewood-1959.1}--\cite{Littlewood-1959.2},
\cite{Nekhoroshev-1977}--\cite{Nekhoroshev-1979} and section~4
of~\cite{Gio-Loc-2006} for a brief summary of such a kind of results).

The approach based on suitable estimates for Birkhoff normal forms has
allowed to introduce the concept of effective stability for physical
systems, that can be stated as follows. Let us focus on the motions
starting from initial conditions staying in a ball of radius $\rho_0$,
that is centered at an equilibrium point; we say that it is {\it
  effectively stable}, if we can ensure that the solutions of the
Hamilton equations stay at a distance not larger than $\rho>\rho_0$
for an interval of time $T$ that is exceeding the (expected) life-time
of that system. This kind of approach has been successfully applied to
the so called Circular Planar Restricted Three-Body Problem
(hereafter, CPRTBP), where the vicinity of the triangular Lagrangian
points ($L_4\,$ and $L_5$) in a model including Sun and Jupiter as
primary bodies is considered. In~\cite{Giorgilli-1997}, it is shown
that the orbits of the third massless body stay within a ball
$B_{\rho}(\vet 0)$ having radius $\rho$ and centered in $L_4$ or $L_5$
for times larger than the age of the universe, where the value of
$\rho$ is approximately the same as $\rho_0$ and the ball
$B_{\rho_0}(\vet 0)$ is large enough to cover the initial conditions
corresponding to the observations of four Trojan Asteroids. Such a
result has been extended in several ways, with applications to
problems arising, mainly, in Celestial Mechanics; the following list
is far from being exhaustive.  In~\cite{Sko-Dok-2001}, a Trojan
Asteroid is shown to be effectively stable in the 3D extension of the
model described just above.  In~\cite{Gab-Jor-2001}, the effects
induced by Saturn are taken into account indirectly, because the
problem is restricted in such a way that the orbit of Jupiter is
prescribed so as to be both periodic and a good approximation of a
numerical solution for a model including Sun, Jupiter and Saturn. In
the models studied in~\cite{Eft-San-2005} and~\cite{Lho-Eft-Dvo-2008},
the continuous Hamiltonian flows have been replaced by symplectic
mappings describing a CPRTBP and an Elliptic Restricted Three-Body
Problem, respectively.  In~\cite{San-Lho-Lem-2014}, the effective
stability of the rotational motion of Mercury is shown in the
framework of a Hamiltonian model.  Last but not least,
in~\cite{San-Gio-Car-2016} Birkhoff normal forms are used in order to
efficiently calculate counter-terms, with the purpose of controlling
the diffusion in a symplectic map of H\'enon type. This has been made
also in view of further possible extensions to similar models
describing the dynamics of particle accelerators.

In spite of its paramount importance, in none of the previously
mentioned articles the construction of the Birkhoff normal form has
been complemented with any {\it rigorous} computer-assisted proof. On the
one hand, the results obtained in all these papers are based on the
explicit calculation of huge numbers of coefficients appearing in the
Birkhoff normal forms, made by using software packages that are
suitable for performing algebraic manipulations. On the other hand,
the computations do not implement interval arithmetic; moreover, for
what concerns the series expansions appearing everywhere in the
algorithms, the analytic estimates of both their radii of convergence
and the sup-norms are missing. The numerical evaluation of such
quantities is often made by using simplistic arguments. In this sense,
those results are not {\it rigorous} proofs. However, completely
rigorous computer-assisted estimates have been carried out in a paper
that is focused on CPRTBP and is an ancestor of a few of those
mentioned above, but its results were strongly limited by a too naive
choice of the initial Hamiltonian expansions
(compare~\cite{Cel-Gio-1991} with~\cite{Giorgilli-1997}).

Computer-assisted proofs have been used in several different fields of
Mathematics, for instance, ranging from combinatorial theory (the
celebrated Four-Colour theorem,
see~\cite{App-Hak-1977}--\cite{App-Hak-Koc-1977}) to PDEs (for a
recent work, see, e.g.,~\cite{Balasz-et-al-2018}).  In particular,
they have allowed to increase very remarkably the threshold of
applicability with respect to the small parameter, that is a crucial
problem in KAM theory (see~\cite{Kolmogorov-1954}, \cite{Moser-1962}
and~\cite{Arnold-1963}).  In fact, while at the very beginning the
first purely analytical results were so limited that they were
completely inadequate for any realistic application, recent rigorous
computer-assisted estimates have been able to get very close to the
optimal breakdown threshold on the small parameter, at least for some
specific problems (see~\cite{Henon-1966} and~\cite{Fig-Har-Luq-2017},
respectively). Until now, the performances of approaches based on the
application of KAM theorem are significantly better with respect to
those using Birkhoff normal forms (for instance,
compare~\cite{Gio-Loc-San-2017} with~\cite{San-Loc-Gio-2013}, where a
secular planetary model including Sun, Jupiter, Saturn and Uranus is
considered).  For what concerns the very specific case of the CPRTBP,
\cite{Gab-Jor-Loc-2005} includes rigorous computer-assisted proofs of
existence of KAM tori in the vicinity of three Trojan
Asteroids. Moreover, in that same work it is shown the numerical
evidence that the algorithm constructing the Kolmogorov normal form is
successful in 23 cases over the first 34 Trojan Asteroids listed in
the IAU Catalogue. Let us recall that more than $7\,000$ asteroids
have been detected about the triangular Lagrangian points of the
system having Sun and Jupiter as primary bodies. Such a large number
of observed objects, compared with the rather small one for which the
previously described results are available, highlights the need for
improvements of the mathematical framework, in order to fully explain
the orbital stability of these celestial bodies.

Our work aims to contribute in filling some of the gaps that have been
discussed above.  In particular, we want to settle a completely
rigorous scheme of computer-assisted estimates for Birkhoff normal
forms.  Moreover, we want to test its performances. In view of such a
purpose, we decided to focus just on a pair of Hamiltonian systems,
that are fundamental examples and have two degrees of freedom.  First,
we apply our method to the H\'enon-Heiles model
(see~\cite{Hen-Hei-1964} and the very recent review
in~\cite{Contopoulos-2020}), that is a very simple example of a
perturbed couple of harmonic oscillators. In particular, we study the
case with two non-resonant frequencies having opposite signs, because
this implies the failure of any trivial proof scheme of the stability.
Indeed, the Hamiltonian is not a Lyapunov function because of the lack
of convexity around the origin.  As a second application, we
reconsider the classical problem of stability of the triangular
Lagrangian equilibria in the CPRTBP.  Let us recall that the structure
of the Hamiltonian for the latter model is very similar to that of the
former one; in particular, trivial approaches to the problem of
stability are hopeless for both. For what concerns the CPRTBP, we
study a variety of situations by considering a few sub-cases where the
role of the second body is played by some different planets. This
allows us to investigate the problem also with very small mass ratios
between the primaries and to check when an approach based on {\it
  resonant} Birkhoff normal forms can be advantageous. Let us recall
that the first results about the expansions of the formal integrals of
motion for resonant systems go back at the dawn of the computer age
and refer to the H\'enon-Heiles model (see,
e.g.,~\cite{Gustavson-1966}).

While in KAM theory there are powerful computer-assisted techniques
that are not based on the Hamiltonian formalism for canonical
transformations (see~\cite{Cel-Chi-2007} and~\cite{Fig-Har-Luq-2017}),
we emphasize that the results about effective stability are
intrinsically related with normal forms. In order to explicitly
construct them, we use the composition of the Lie series, that is a
very mature algorithm and can be complemented with a complete scheme
of analytical estimates (see~\cite{Giorgilli-2003.1} for an
introduction).

The paper is organised as follows. In section~\ref{sec:birk}, we
describe the construction of the Birkhoff normal form by an approach
based on Lie series.  This is the starting point for the definition of
the scheme of estimates, which is explained in
section~\ref{sec:stime}. In section~\ref{sec:eff-stab}, we discuss how
to use the estimates to prove effective stability for Birkhoff normal
forms in both the non-resonant case and the resonant one. In
section~\ref{sec:results}, we describe our results for the
H\'enon-Heiles model and for the CPRTBP. Main conclusions are drawn in
section~\ref{sec:conclu}. Appendix~\ref{app:interval} is devoted to an
introduction to those concepts of validated numerics, that are
essential to implement our computer-assisted proofs in a fully
rigorous way. In appendix~\ref{app:example}, we describe in detail an
application to the H\'enon-Heiles model in a situation, that is
tailored in such a way that a relatively small number of computations
are needed to obtain a (very limited) result; this small tutorial
example is discussed with the aim of making our work more easy to
reproduce for a reader that is interested in exporting our approach in
other challenging contexts.

\section{Construction of Birkhoff normal form around an elliptic equilibrium point}
\label{sec:birk}

In this section we are going to describe the construction of the
Birkhoff normal form for a particular class of Hamiltonians. This is
made by using Lie series operators. Although such a method is rather
standard in perturbation theory, a detailed description of the formal
algorithm is mandatory, in order to make well definite the discussion
in the next sections. The adoption of an approach based on Lie series
makes our work further different with respect to those that have been
mentioned in the Introduction and deal with effective stability
estimates based on Birkhoff normal forms, because they implemented
algorithms using Lie transforms.

Let us give some definition.  Let $f$ be a polynomial function in the
canonical variables $(\vet y, \vet x)\in\reali^n\times\reali^n$, then
we define the Lie derivative operator with respect to $f$ as $\lie f
(\cdot) = \poisson f \cdot$, being $\poisson \cdot \cdot$ the Poisson
bracket\footnote{Let us recall the definition of the Poisson bracket:
  $\poisson f g = \sum_{j=1}^n \left(\frac{\partial f}{\partial x_j}
  \frac{\partial g}{\partial y_j} - \frac{\partial f}{\partial y_j}
  \frac{\partial g}{\partial x_j}\right)$.} between two
functions. Moreover, the Lie series operator having $f$ as generating
function is defined as follows:
\begin{equation}
\label{frm:lie-serie}
 \exp \lie {f} (\cdot)  = \sum_{j=0}^{+\infty}\frac{1}{j!}\Lie{f}^j (\cdot) \,
\end{equation}
and it will be used to define near to the identity canonical
transformations.  Moreover, we denote by $\Pscr_s$ the class of
homogeneous polynomials of degree $s$ in the canonical variables
$(\vet y, \vet x)$.  In the following lemma, we describe the behaviour
of the homogeneous polynomials with respect to the Poisson brackets.
\begin{lemma}
\label{lem:poisson-classi}
Let $f$ and $g$ be polynomial functions in $\Pscr_{r+2}$ and
$\Pscr_{s+2}\,$, respectively. Therefore, $\poisson f g \in
\Pscr_{s+r+2}\,$. Moreover, $\Lie {f}^j g \in \Pscr_{jr+s+2} \ \forall
\ j \ge 0\,$.
\end{lemma}
This simple property is fundamental, because it will be repeatedly
used for reorganizing the Hamiltonian terms (appearing in an
expansion) with respect to their polynomial degree, after each
canonical change of coordinates defined by a Lie series operator.

\subsection{Hamiltonian framework}
Since we aim to study the dynamics in the neighbourhood of an elliptic
equilibrium point, located in correspondence with the origin, we
consider a Hamiltonian of the following type:
\begin{equation}
\label{frm:h0}
\Hscr(\vet y,\vet x)=
\sum_{j=1}^n \omega_j \frac{y_j^2+x_j^2}{2}+
\sum_{\ell=1}^{+\infty} f_\ell(\vet y,\vet x)
\, \quad {\rm with } \quad f_\ell \in \Pscr_{\ell+2},\ \vet \omega\in \reali^n.
\end{equation}
In the proximity of the origin, the magnitude order of the
polynomial terms $f_\ell$ is obviously smaller than that of the main
part of the Hamiltonian, which is given by a sum of harmonic
oscillators. Moreover, if we introduce the action-angle coordinates
for harmonic oscillators, by the following canonical change of
variables
\begin{equation}
  \label{frm:da_varcanpol_a_azang}
  x_j = - \sqrt{2I_j}\cos{(\phi_j)}
  \qquad
  y_j = \sqrt{2I_j}\sin{(\phi_j)}
  \qquad \forall \ j  =1, \ldots, n,
\end{equation}
then the Hamiltonian in~\eqref{frm:h0} transforms to
\begin{equation}
\label{frm:h0-iphi}
\Hscr(\vet I, \vet \phi)=
\vet \omega \cdot \vet I +
\sum_{\ell=1}^{+\infty} f_\ell(\vet I, \vet \phi)
\quad {\rm with}\quad  (\vet I, \vet \phi)\in \reali^n \times \toro^n\,,
\end{equation}
with $f_\ell$ containing terms having total degree in the square
root of the actions $I_1\,,\,\ldots\,,\,I_n$ equal to $\ell$ and
trigonometric expansions in $\vet \phi$ with Fourier harmonics $\vet
k$ such that $|\vet k|=\sum_j|k_j|\le \ell$.  These variables
highlight the integrability of the quadratic part, which depends on
the actions only. Therefore, we will proceed in a perturbative way,
by leading the Hamiltonian to a Birkhoff normal form. This means that,
after having performed $r$ canonical changes of coordinates
defined by the corresponding Lie series, the Hamiltonian will be such
that
\begin{equation}
  \Hscr^{(r)}(\vet I, \vet \phi)=
  \Zscr^{(r)}(\vet I) + \Rscr^{(r)}(\vet I, \vet \phi) \,,
\label{frm:hr}
\end{equation}
where the upper index $r$ in every Hamiltonian term refers to the
number of normalization steps that have been already performed.
In some more words, we want to remove step by step the angular
dependence from the perturbative part of the Hamiltonian, in order to
decrease the size of the remainder $\Rscr^{(r)}$. This is done to
provide a better integrable approximation $\Zscr^{(r)}$ with respect
to the problem we are studying.

In order to deal with the construction of the normal form and the
subsequent scheme of estimates, it is convenient to work with the
complex canonical variables $(-i\vet z, \bar{\vet z})$, introduced by
the following canonical transformation
\begin{equation}
z_j = -\sqrt{I_j}e^{-i\phi_j} \quad \forall \ j = 1, \ldots, n \,.
\label{frm:variabili-complesse}
\end{equation}
In these new variables $I_j = z_j \bar z_j\,$, then a new normal form
term should contain only monomials where $z_j$ and $\bar z_j$ appear
with the same exponent. Moreover, since the polynomial degree is the
same if we are using either the canonical variables $(\vet y, \vet x)$
or the complex ones $(-i \vet z, \bar{\vet z})$, we will maintain the
symbol $\Pscr_{\ell}\,$, $\forall\ \ell\ge 0$, to denote every class
of homogeneous polynomials.

\subsection{Description of the $r$-th normalization step}

We are going to describe in detail how to perform a single step of the
normalization algorithm.  Let us suppose that we have been able to
perform the first $r-1$ steps, in such a way that the Hamiltonian is
given the following form:
\begin{equation}
  \Hscr^{(r-1)}(-i\vet z, \bar{\vet z})=
  \sum_{\ell=0}^{r-1}Z_\ell(\vet z  \bar{\vet z})
  + \sum_{\ell=r}^{+\infty} f_\ell^{(r-1)}(-i\vet z, \bar{\vet z}) \,,
\label{frm:hr-1}
\end{equation}
where $Z_\ell \in \Pscr_{\ell+2}$ is depending on the actions only and
$f_\ell^{(r-1)}\in \Pscr_{\ell +2}\,$, while the upper index $r-1$
(appearing also in the symbol denoting the Hamiltonian) refers to the
normalization step. When $r=1$, this is nothing but
Hamiltonian~\eqref{frm:h0-iphi} once it has been expressed in complex
variables. We want to normalize the main term among the perturbative
ones, that is $f_r^{(r-1)}$.  For this purpose, we introduce the new
$r$-th Hamiltonian as follows\footnote{In
  formula~\eqref{frm:da-hr-1-a-hr}, we make use of the so called {\it
    exchange theorem} (see, e.g., formula~(4.6)
  in~\cite{Giorgilli-2003.1}) which states that, if $\chi$ is a
  generating function, $f({\vet p},{\vet q})|_{({\vet p},{\vet q}) =
    \exp \Lie \chi ({\vet p}',{\vet q}')} = \exp \Lie \chi f|_{({\vet
      p},{\vet q}) = ({\vet p}',{\vet q}')}$. Therefore, we can apply
  the Lie series to the Hamiltonian function and, only at the end,
  rename the variables. For more detailed explanations we defer to the
  whole section~4.1 of~\cite{Giorgilli-2003.1}. Here, we do not rename
  the variables in order to avoid the proliferation of too many
  symbols. }:
\begin{equation}
\label{frm:da-hr-1-a-hr}
\Hscr^{(r)} (-i\vet z, \bar{\vet z}) =
\exp \lie {\chi_r} \Hscr^{(r-1)} (-i\vet z, \bar{\vet z}) =
\sum_{j=0}^{+\infty}\frac{1}{j!}\Lie{\chi_r}^j  \Hscr^{(r-1)}(-i\vet z, \bar{\vet z})\,.
\end{equation}
Since we are asking that the term of polynomial degree equal to $r+2$
is depending just on the actions, then we have to determine the
generating function $\chi_r$ in such a way that the following
homological equation is satisfied:
\begin{equation}
\Lie {\chi_{r}} Z_{0}+ f_{r}^{(r-1)}= Z_{r} \,,
\label{frm:eq-omologica}
\end{equation}
where $Z_0$ is the quadratic part, i.e., $Z_0 = \sum_{j=1}^n \omega_j
z_j \bar z_j$, and $Z_r$ is the new term in normal form we have to
define. Hereafter, it is convenient to adopt the standard multi-index
notation; for instance, this means that $(- i\vet z
)^{\vet\ell}=\prod_{j=1}^{n} (-i\,z_j)^{\ell_j}$, $\forall\ \vet z \in
\complessi^n$ and $\vet \ell \in \naturali^n$. Moreover, for what
concerns the complex conjugate variables, the notation has to be read
as follows: $ \bar{\vet z}^{\tilde{\vet\ell}}=\prod_{j=1}^{n} {\bar
  z}_j^{\tilde{\ell}_j}$, $\forall\ \vet z \in \complessi^n$ and
$\tilde{\vet\ell} \in \naturali^n$.  Let us write the generic
expansion of the perturbative term $f_r^{(r-1)}$ so that
\begin{equation}
\label{frm:f-r}
f_r^{(r-1)} = \sum_{|\vet \ell|+|\tilde{\vet \ell}| = r+2}
c_{\vet \ell, \tilde{\vet \ell}}^{(r-1)}  \bar{\vet z}^{\tilde{\vet\ell}}(- i\vet z )^{\vet\ell}
\qquad {\rm with} \qquad c_{\vet \ell, \tilde{\vet \ell}}^{(r-1)}
\in \complessi\,,\ \vet\ell, \tilde{\vet \ell} \in \naturali^n;
\end{equation}
therefore, the generating function that solves the homological
equation~\eqref{frm:eq-omologica} is such that
\begin{equation}
\label{frm:chi-r}
\chi_r = \sum_{|\vet \ell|+|\tilde{\vet \ell}| = r+2 \atop{\vet \ell \neq \tilde{\vet \ell}}}
-\frac{c_{\vet \ell, \tilde{\vet \ell}}^{(r-1)}}
{i \vet \omega \cdot (\vet \ell- \tilde{\vet \ell})}
\bar{\vet z}^{\tilde{\vet\ell}}(- i\vet z )^{\vet\ell}\,,
\end{equation}
where $\vet \omega$ is the vector of frequencies of the unperturbed
harmonic oscillators in~\eqref{frm:h0}.  Clearly, this generating
function can be properly defined if and only if the frequency vector
$\vet \omega$ is non-resonant, otherwise $\vet \omega \cdot (\vet
\ell- \tilde{\vet \ell})$ could vanish.  Therefore, we assume that $\vet
\omega$ satisfies the Diophantine inequality, that is
\begin{equation}
\label{frm:non-res}
|\vet k \cdot \vet \omega| \ge \frac{\gamma}{|\vet k|^\tau} \quad
\forall \ \vet k \in \interi^n \setminus \{ \vet 0 \}\,,
\end{equation}
for some fixed values of $\gamma>0$ and $\tau \ge n-1$. This condition
is not so strict, since for $\tau >n-1$ it is satisfied by almost all
the frequency vectors in $\reali^n$. The terms with $\vet \ell =
\tilde{\vet \ell }$, that cannot be removed by this procedure, depend
on the actions only, since $\vet z$ and $\bar{\vet z}$ appear with the
same exponent.  As a consequence, we are forced to include them in the
new term $Z_r$ making part of the normal form, that is
\begin{equation}
\label{frm:Z_r}
Z_r = \sum_{2|\vet \ell| = r+2 } c_{\vet \ell, \vet \ell}^{(r-1)}
(-i \vet z \bar{\vet z})^{\vet\ell}\,.
\end{equation}
From definitions~\eqref{frm:chi-r} and~\eqref{frm:Z_r}, it is evident
that both $\chi_r$ and $Z_r$ belong to $\Pscr_{r+2}\,$.

In order to conclude the normalization step, we have to perform the
canonical change of coordinates induced by $\exp \Lie {\chi_r}$ and to
update accordingly all the terms that compose the Hamiltonian. For
this purpose, it is convenient to use a notation which mimics a
programming code. At first, we put the new terms
$f_\ell^{(r)}=f_\ell^{(r-1)}$; this simply corresponds to the initial
summand with $j=0$ in~\eqref{frm:lie-serie}. Then, each contribution
due to the $j$-th iteration of the Lie derivative with respect to the
generating function $\chi_r$ (for $j\ge 1$) is added to the
corresponding class. More precisely, using repeatedly the property
described in lemma~\ref{lem:poisson-classi}, the new summands, which
are generated by the application of the Lie series to the terms in
normal form (that are denoted with $Z_s$) and to the perturbative ones
(i.e., $f_s^{(r-1)}$), are gathered in the following way:
\begin{equation}
\vcenter{\openup1\jot\halign{
 \hbox {\hfil $\displaystyle {#}$}
&\hbox {\hfil $\displaystyle {#}$\hfil}
&\hbox {$\displaystyle {#}$\hfil}\cr
f_{s+jr}^{(r)} & \hookleftarrow
& \frac{1}{j!}\Lie {\chi_r}^j Z_s \phantom{01234}
\forall \ 0 \le s < r, \ j \ge 1 \, ,
\cr
f_{s+jr}^{(r)} & \hookleftarrow & 
\frac {1}{j!} \Lie {\chi_r}^j f_s^{(r-1)}
\quad \forall \  s \ge r, \ j \ge 1 \, ,
\cr
}}
\label{frm:nuovi-pert}
\end{equation}
where the notation $a\hookleftarrow b$ means that $a$ is redefined
so as to be equal to the sum given by its previous value plus $b$.
At the end of all these redefinitions, the following expansion of the
new Hamiltonian is well defined:
\begin{equation}
  \Hscr^{(r)}(-i\vet z, \bar{\vet z})=
  \sum_{\ell=0}^{r}Z_\ell(\vet z  \bar{\vet z})
  + \sum_{\ell=r+1}^{+\infty} f_\ell^{(r)}(-i\vet z, \bar{\vet z}) \,.
\label{frm:hr-expansion}
\end{equation}

Let us emphasize that in this procedure we do not modify the terms
that are already in normal form. This is why the integrable part has
nearly the same expression when~\eqref{frm:hr-1} is compared
with~\eqref{frm:hr-expansion}, with just one exception due to the
occurrence of a new normal form term $Z_r\,$, while the perturbative
terms are different, being redefined as it has been prescribed by
formula~\eqref{frm:nuovi-pert}.

The algorithm can be iterated at the next step, by restarting from the
Hamiltonian~\eqref{frm:hr-1} where $r-1$ has to be replaced with $r$.

\subsection{About the divergence of the Birkhoff normal form}

In spite of the fact that we are interested in providing results
holding true for a ball of initial conditions $B_{\rho_0}(\vet 0)$,
where $\rho_0$ is {\it fixed}, now we are going to briefly discuss
some asymptotic properties of the Birkhoff normal form.

It is well known that the series introduced by the algorithm
constructing the normal form are asymptotically divergent, with the
exception of some particular cases. This means that the sup-norm of
the remainder $\Rscr^{(r)}$ (appearing in~\eqref{frm:hr}) does not go
to zero for $r$ which tends to infinity. However, the Birkhoff normal
form is still very useful, because there is an optimal normalization
step which minimizes the remainder.  The mechanism of divergence is
mainly due to the accumulation of the so called ``small divisors".  In
fact, at each step $r$, new divisors $\vet \omega \cdot (\vet \ell-
\tilde{\vet \ell})$ (being $|\vet\ell| + |\tilde{\vet \ell}| = r+2$) are
introduced by the definition of the generating function $\chi_r$
in~\eqref{frm:chi-r}. Because of the redefinitions described in
formula~\eqref{frm:nuovi-pert}, the size of the generating functions
clearly impacts on the growth of the terms constituting the
remainder. If the frequency vector satisfies the Diophantine condition
in~\eqref{frm:non-res}, it is rather easy to see that factorial
coefficients $\Oscr(r!)$ appear in the estimate of the remainder at
the $r$-th normalization step. This prevents the convergence of the
normalization algorithm when it is iterated {\it ad infinitum}.

Nevertheless, when the series are estimated on open balls
$B_{\rho}(\vet 0)$, with a fixed value of $\rho$, it is convenient to
proceed with the normalization algorithm until the remainder is
decreasing. The optimal step $r_{\rm opt}$ is determined, by comparing
two subsequent remainders $\Rscr^{(r)}$ and $\Rscr^{(r-1)}$. As
discussed, e.g., in~\cite{Giorgilli-2003.1} (see formula~(3.30)), this
allows to conclude that
\begin{equation}
\label{frm:passo-ottimale}
 r_{\rm opt} \sim ( C\rho ) ^{-{\frac{1}{\tau + 1}}} \, ,
\end{equation}
where $C$ is a positive constant, $\tau$ is the exponent appearing in
the Diophantine condition~\eqref{frm:non-res} and $\rho$ is the
distance from the equilibrium point. As a consequence, an analytical
estimate of the remainder at the optimal step can be easily provided;
in fact, it is exponentially small with respect to the inverse of a
fractional power of the distance from the equilibrium $\rho$, that in
this case assumes the role of small parameter.  Near the elliptic
equilibrium, the Birkhoff normal form is a good approximation of the
real problem and it is convenient to perform a big number of
normalization steps. On the other hand, far away from the origin the
Birkhoff normal form starts to diverge earlier. Let us mention that
in~\cite{Eft-Gio-Con-2004} the mechanism of divergence is carefully
investigated in a numerical way and it is shown to be much more subtle
with respect to what has been discussed just above. Moreover, this has
allowed those authors to conclude that the analytical upper bounds
should largely overestimate the effective size of the remainders.

In practice, when dealing with explicit expansions, two additional
problems have to be tackled: truncations and computational costs. In
fact, the Hamiltonian at step $r$ is expanded as in~\eqref{frm:hr}
and~\eqref{frm:hr-expansion}, with
\begin{equation}
\label{frm:resto}
\Rscr^{(r)} = f_{r+1}^{(r)} +  f_{r+2}^{(r)} + \ldots
\quad {\rm with} \quad f_{\ell}\in \Pscr_{\ell+2}\ \ \forall \ \ell >r.
\end{equation}
Of course, efficient coding and a powerful computing system are more
than welcome, because they allow better evaluations of the remainder
terms. However, any computer cannot represent all the infinite
sequence of terms giving contributions to $\Rscr^{(r)}$.  In the next
section, we will explicitly provide rigorous upper bounds for the
series defining the remainder.

\section{Iterative estimates for Birkhoff normal form}
\label{sec:stime}

In this section we are going to discuss how to construct a scheme of estimates, in order to keep control of the norms of the
Hamiltonian terms, starting from the algorithm described in
section~\ref{sec:birk}. In spite of the fact that every purely
analytical work based on Birkhoff normal forms exploits a suitable
scheme of estimates, our approach significantly differs from all the
previously existing ones in the scientific literature for what concerns
the following untrivial point, that is clearly highlighted, for
instance, in the statements of
propositions~\ref{pro:itera-stime-singole}
and~\ref{pro:itera-stime-code}. Here, the estimates are designed in
such an iterative way, that they take advantage of being applied in the
framework of a computer-assisted proof, in order to provide the best
possible final results.

Before proceeding with the description of the scheme of estimates, the
introduction of some new notation is mandatory. For every $f \in
\Pscr_s$ whose expansion is the following:
\begin{equation}
  f= \sum_{|\vet \ell|+ |\tilde {\vet \ell}| = s} c_{\vet \ell, \tilde {\vet \ell} }
  \, (-i \vet z)^{\vet \ell}\bar {\vet z}^{\tilde {\vet \ell}} \,,
\end{equation}
we denote by $\| \cdot \|$ the functional norm defined as
\begin{equation}
\label{frm:norma}
\| f \|= \sum_{|\vet \ell|+ |\tilde {\vet \ell}| = s}
|c_{\vet\ell, \tilde{\vet \ell}}|  \, .
\end{equation}
Therefore, in the domain 
$$
\Delta_{\rho}=
\{(-i \vet z, \bar {\vet z}):\ |z_j| < \rho \,, \ 1 \le j \le n\} =
\{ (\vet I, \vet \phi):\ I_j < \rho^2 \,,\ \phi_j\in\toro\,,\ 1 \le j \le n\}\,,
$$
the $\sup$ norm of $f$ can be easily estimated as follows:
\begin{equation}
\label{frm:stima-norma-sup}
|f|_{\rho} = \sup_{\Delta_{\rho}} |f(-i\vet z, \bar{\vet  z})|
\le \rho^{s}  \| f\| \,.
\end{equation}

For the sake of simplicity, in the previous definitions we have not
introduced more general domains that are polydisks; this has been done
with the aim to avoid all the modifications that are necessary to adapt norms
and estimates.  However, let us recall that in some applications of
the Birkhoff normal forms the approach based on polydisks has been
useful (see, e.g.,~\cite{Giorgilli-1997}).

Hereafter, we will assume to have explicitly built the Birkhoff normal
form up to a fixed order $r = R_{\rm I}$ and we will define estimates
for the norms of the infinite terms we did not calculate. In order to
do that, for every normalization step we will distinguish between two
classes of terms making part of the power series expansion: those
belonging to $\Pscr_\ell$ with $\ell \le R_{\rm II}$, for which we
will give explicit estimates of the norms, and the infinite terms with
degree greater than $R_{\rm II}+2$, that we will control by a sequence
of majorants growing in a geometrical way.  In the next subsections,
we describe how to define iteratively the estimates for the norms of
these two kinds of polynomial terms.

\subsection{Iterative estimates for finite expansions of the power series}

Let us suppose that we have already performed the first $r-1$ steps of
normalization, so that our Hamiltonian is expanded as
in~\eqref{frm:hr-1}; moreover, let us assume to know upper bounds for
the norm of every term appearing in~\eqref{frm:hr-1}, i.e.,
\begin{equation}
\label{frm:stime-norme-termini}
\vcenter{\openup1\jot\halign{
 \hbox {\hfil $\displaystyle {#}$}
&\hbox {\hfil $\displaystyle {#}$\hfil}
&\hbox {$\displaystyle {#}$\hfil}\cr
\| Z_s\| &\le \Zscr_s  \quad &  \forall \ 0 \le s \le r-1\,, \cr
\| f_s^{(r-1)}\|  &\le \Fscr_s^{(r-1)}  \quad  &\forall \ s \ge r\,,
\cr
}}
\end{equation}
with $\Zscr_s, \ \Fscr_s^{(r-1)} \in \reali^+$. We have now to
describe how to estimate the norms of the terms that compose the new
Hamiltonian $\Hscr^{(r)}$, namely how the estimates $\Fscr_s^{(r-1)}$
change after a step of Birkhoff normalization.  Since the canonical
change of coordinates that transforms $\Hscr^{(r-1)}$ into
$\Hscr^{(r)}$ is defined by a Lie series with generating function
$\chi_r$, we need an estimate for the norm of the Poisson brackets
between two functions. In the following lemma, we adapt to the present
context a well known estimate that can be found, e.g.,
in~\cite{Gio-Del-Gal-Sim-1989}.
\begin{lemma}
Let $f$ and $g$ be polynomial functions such that $f \in \Pscr_{s+2}$,
$g \in \Pscr_{r+2}$ and suppose that $\|f\| \le \Fscr$ and $\| g\| \le
\Gscr $ for some constant $\Gscr,\Fscr \in \reali^+$; therefore,
\begin{equation} 
\label{frm:stima-poisson}
\|\poisson f g \| \le (s+2)(r+2) \Fscr \Gscr\,.
\end{equation}
\label{lem:stima-poisson}
\end{lemma}
In the framework of computer-assisted estimates, sometimes it is more
convenient to use the following lemma.
\begin{lemma}
\label{lem:stima-lie}
Let $\chi_r$ and $g$ be polynomial functions with $\chi_r \in
\Pscr_{r+2}$ and $g \in \Pscr_{s+2}\,$, being the expansion of the
former function such that
\begin{equation}
\label{frm:chi}
\chi_r = \sum_{|\vet k|+|\tilde{\vet k}| = r+2} c_{\vet k, \tilde{\vet k}}
(- i\vet z )^{\vet k}  \bar{\vet z}^{\tilde{\vet k}}\,.
\end{equation}
Therefore, $\forall \ j \ge 1$, the following estimate holds true:
\begin{equation}
\label{frm:stima-lie}
\left\|  \frac{1}{j!} \Lie {\chi_r}^j g \right\| \le
\frac{\prod_{i=0}^{j-1} (s+ir+2)}{j!} \Dscr_r^j \|g \|  \, ,
\end{equation}
where $\Dscr_r = \sum_{|\vet k|+|\tilde{\vet k}| = r+2} 
|c_{\vet k, \tilde{ \vet k}}| \max_j \{|k_j|, |\tilde{ k_j}|\}$.
\end{lemma}
Indeed, if we replace $\Dscr_r$ with $(r+2)\|\chi_r\|$, the previous
statement easily follows by induction from
lemma~\ref{lem:stima-poisson}. The above definition of $\Dscr_r$
allows to estimate more carefully the partial derivatives of $\chi_r$
that appear in the Poisson brackets. This is particularly advantageous
when the optimal normalization step is not greater than the maximal
value $R_{\rm I}$ of the index $r$ for which we compute explicitly the
expansion of the generating function $\chi_r$. On the other hand, when
the number $R_{\rm I}$ is lower than the optimal step $r_{\rm opt}\,$, it
is convenient to provide an estimate for the norm $\| \chi_r\|\le
\Gscr_r$ and, then, $\Dscr_r$ can be replaced by $(r+2)\Gscr_r\,$, for
all those $r\in [R_{\rm I}+1\,,\,r_{\rm opt}]$ for which the coefficients
$c_{\vet k, \tilde{\vet k}}$ appearing in the expansion~\eqref{frm:chi}
are unknown.

In order to determine upper bounds for the norms of the new
perturbative terms (introduced by the $r$-th normalization step of the
algorithm), we have to focus on the redefinitions in
formula~\eqref{frm:nuovi-pert} for using
inequality~\eqref{frm:stima-lie}. Therefore, it is easy to realize
that we substantially need to find an estimate for the norm of the
generating function. Obviously, if $r\le R_{\rm I}$, we know exactly
the expression of $\chi_r$ and we can calculate $\Dscr_r$ as
prescribed in lemma~\ref{lem:stima-lie}. When the normalization step
$r>R_{\rm I}\,$, recalling the expansion of $\chi_r$
in~\eqref{frm:chi-r}, we can say that the coefficients of the
generating function are equal to those of $f_r^{(r-1)}$ divided by
$\vet\omega\cdot (\vet \ell-\tilde{\vet \ell})$. Let $\alpha_r$ be the
smallest divisor introduced at the $r$-th normalization step, i.e.,
\begin{equation}
\label{frm:alfa-r}
\alpha_r = \min_{|\vet\ell| + |\tilde{\vet \ell}| = r+2 \atop{\vet \ell \neq \tilde{\vet \ell}}}
|\vet\omega\cdot (\vet \ell-\tilde{\vet \ell})|\,,
\end{equation}
which is well-defined in view of the non-resonance condition assumed
in~\eqref{frm:non-res}. Therefore, we can write $\|\chi_r\|\le\Gscr_r\,$,
with
\begin{equation}
\label{frm:g-r}
\Gscr_r = \frac{\Fscr_r^{(r-1)}}{\alpha_r}\,.
\end{equation}
Analogously, by definition of the new term in normal form $Z_r$
in~\eqref{frm:Z_r}, the norm of $Z_r$ cannot be greater than $\Zscr_r
= \Fscr_r^{(r-1)}$.  We are now ready to define the new majorants
$\Fscr_s^{(r)}$, providing upper bounds for the perturbative terms
appearing in the Hamiltonian $\Hscr^{(r)}$.
\begin{proposition}
\label{pro:itera-stime-singole}
Let us suppose that the terms of the Hamiltonian introduced at
$r-1$-th normalization step are bounded as
in~\eqref{frm:stime-norme-termini}. Therefore, the new Hamiltonian
terms of $\Hscr^{(r)}$ are bounded so that $\| Z_r\| \le \Zscr_r =
\Fscr_r^{(r-1)}$ and $\| f_s^{(r)}\|\le\Fscr_s^{(r)}$ $\forall \ s >
r$, where these new constants are given by the following sequence
of redefinitions:
\begin{equation}
\label{frm:stime-ridef}
\vcenter{\openup1\jot\halign{
 \hbox {\hfil $\displaystyle {#}$}
&\hbox {\hfil $\displaystyle {#}$\hfil}
&\hbox {$\displaystyle {#}$\hfil}\cr
\Fscr_\ell^{(r)} & = \Fscr_\ell^{(r-1)} & \ \forall \ \ell \ge 0 \, ,
\cr
\Fscr_{s+jr}^{(r)} & \hookleftarrow
\frac{ \prod_{i=0}^{j-1} (s+ir+2)}{j!} \Dscr_r^j \Fscr_s^{(r-1)}
& \ \forall \ j \ge 1, \  s \ge 0 \, ,
\cr
}}
\end{equation}
being $\Dscr_r = \sum_{|\vet k|+|\tilde{\vet k}| = r+2} |c_{\vet k, \tilde{ \vet
    k}}| \max_j \{|k_j|, |\tilde{ k_j}|\}$, if the expansion of the
generating function $\chi_r$ is explicitly known, as it is written
in~\eqref{frm:chi}, else $\Dscr_r =(r+2)\Gscr_r\,$, with $\Gscr_r =
{\Fscr_r^{(r-1)}}/{\alpha_r}\,$.
\end{proposition}
The previous statement is basically a summary of all the discussion
explained in the present subsection. In particular,
formula~\eqref{frm:stime-ridef} can be easily verified, starting from
the iterative definitions of the new perturbative Hamiltonian terms
in~\eqref{frm:nuovi-pert} and using the estimates in
lemma~\ref{lem:stima-lie}.

\subsection{Power series expansions of Birkhoff normal forms: geometrical increase of majorants for the infinite tails of terms}

In this subsection, we are going to describe how to estimate the norm
of the terms that cannot be explicitly calculated or iteratively
estimated, as it has been described in the previous subsection. In
order to do that, it is convenient to bound in a different way the
terms constituting the Hamiltonian at the $r-1$-th normalization
step, i.e.,
\begin{equation}
\| Z_s \| \le \Escr\, a_{r-1}^{s+2}  \quad \forall \ 1 \le s \le r-1 \,,
\qquad
\| f_s^{(r-1)} \| \le \Escr \,a_{r-1}^{s+2} \quad \forall \ s \ge r \,,
\label{frm:stima-per-code}
\end{equation}
with $\Escr, a_{r-1}\in \reali^+$. Of course, the estimates above are
expected to be less strict than the ones
in~\eqref{frm:stime-norme-termini}. However, let us imagine to be able to justify the same kind of estimates at the $r$-th
normalization step, holding true for all $f_s^{(r)}$ with $s \ge r+1$,
with a new constant value $a_{r}\,$; therefore, we could control the
sup norm of the remainder (recall equations~\eqref{frm:hr}
and~\eqref{frm:hr-expansion}) as follows:
\begin{equation}
\label{frm:normasup-resto}
\sup_{\Delta_{\rho }} |\Rscr^{(r)}| \le
\sum_{s=r+1}^{R_{\rm II}} \Fscr_s^{(r)}\rho^{s+2} +
\Escr \sum_{s=R_{\rm II}+1}^{+ \infty}(a_r \rho)^{s+2}\,,
\end{equation}
where we include the upper bounds as iteratively defined by
proposition~\ref{pro:itera-stime-singole} when they are available,
while for the terms with polynomial degree greater than $R_{\rm II}+2$
we use the wanted geometric estimates. These
are obtained by iteration of the rule described below.

\begin{proposition}
\label{pro:itera-stime-code}
Let us assume that the estimates in~\eqref{frm:stima-per-code} hold
true for all the terms appearing in the expansion~\eqref{frm:hr-1} of
$\Hscr^{(r-1)}$.  Therefore, the norms of the terms making part of the
new Hamiltonian $\Hscr^{(r)}$ satisfy the following inequalities:
\begin{equation}
\label{frm:stima-per-code-r}
\| Z_s \| \le \Escr \, a_r^{s+2} \quad \forall \ 1 \le s \le r \, ,
\qquad
\| f_s ^{(r)}\| \le \Escr \, a_r^{s+2} \quad \forall \ s \ge r+1 \, ,
\end{equation} 
being
\begin{equation}
\label{frm:a_r}
a_r= a_{r-1} \left(1+ \frac{ (r+1) \Dscr_r} {a_{r-1}^r}\right)^{\frac 1 r} \,,
\end{equation} 
where $\Dscr_r = \sum_{|\vet k|+|\tilde{\vet k}| = r+2} |c_{\vet k, \tilde{ \vet
    k}}| \max_j \{|k_j|, |\tilde{ k_j}|\}$ if the
expansion~\eqref{frm:chi} of the generating function $\chi_r$ is
explicitly known, else $\Dscr_r= (r+2)\Gscr_r$ with $\| \chi_r \| \le
\Gscr_r\,$.
\end{proposition}

\begin{proof}
Let us start from the estimate for $Z_s$ with $1\le s<r$. Since
$a_{r-1}<a_r\,$, all the estimates in~\eqref{frm:stima-per-code} are
still valid when $a_{r-1}$ is replaced by $a_r\,$. 
By comparing the polynomial expressions~\eqref{frm:f-r} and~\eqref{frm:Z_r} that are
 related to the functions $f_r^{(r-1)}$ and $Z_r\,$, respectively, it
 is easy to realize that the norm of the latter cannot be greater than
 the former one, because in a normal form term there are just a part of
 the summands appearing in expansion of the corresponding perturbative
 function. Therefore, in view of~\eqref{frm:stima-per-code} we can
 write the following chain of inequalities: $\|Z_r\|\le
 \|f_r^{(r-1)}\|\le \Escr\,a_{r-1}^{r+2}\le \Escr\,a_{r}^{r+2}$; this
 ends the justification of the first inequality in formula~\eqref{frm:stima-per-code-r}.

Let us now focus on the estimates for $f_s^{(r)}$, with $s \ge
r+1$. First, we discuss the case with $s =jr+m$ for $m=1, \ldots ,
\ r-1$ and $j \ge 1$: the new perturbative terms are defined as the
sum of $f_s^{(r-1)}$ plus the new contributions due to $\lie{\chi_r}^j
Z_m/j!$ and $\lie{\chi_r}^\ell f_{(j-\ell)r+m}^{(r-1)}/\ell!$ for
$\ell =1, \ldots, \ j-1$.  Therefore, inequality~\eqref{frm:stima-lie}
allows us to write the following estimate:
\begin{equation*}
\vcenter{\openup1\jot\halign{
 \hbox {\hfil $\displaystyle {#}$}
&\hbox {\hfil $\displaystyle {#}$\hfil}
&\hbox {$\displaystyle {#}$\hfil}\cr
\Fscr_{jr+m}^{(r)} &= & \Fscr_{jr+m}^{(r-1)}  +  \frac{1}{j!}\lie{\chi_r}^j Z_m + 
\sum_{\ell=1}^{j-1}  \frac{1}{\ell!} \lie{\chi_r}^\ell f_{(j-\ell)r +m}^{(r-1)} 
\cr
&\le
&\Fscr_{jr+m}^{(r-1)} + \,\frac{\prod_{i=0}^{j-1}(m+ir+2)}{j!} \Dscr_r^j \Zscr_m
\cr
& & + \sum_{\ell=1}^{j-1}
\left( \,\frac{\prod_{i=0}^{\ell-1}(m+(j-\ell+i)r+2)}{\ell!}
\Dscr_r^\ell \Fscr_{(j-\ell)r+m}^{(r-1)} \right) \, .
\cr
}}
\end{equation*}
Using the estimate in hypothesis~\eqref{frm:stima-per-code-r} and the
trivial inequality $m<r$, we obtain
\begin{equation*}
\vcenter{\openup1\jot\halign{
 \hbox {\hfil $\displaystyle {#}$}
&\hbox {\hfil $\displaystyle {#}$\hfil}
&\hbox {$\displaystyle {#}$\hfil}\cr
\Fscr_{jr+m}^{(r)} &\le 
& \Escr\, a_{r-1}^{jr+m+2}+
\, \frac{\prod_{i=0}^{j-1}(i+1)}{j!}[(r+1)\Dscr_r]^j \Escr \,a_{r-1}^{m+2} 
\cr
& & + \sum_{\ell=1}^{j-1}  \,
\frac{\prod_{i=0}^{\ell-1}(j-\ell+i+1)}{\ell! a_{r-1}^{\ell r}}
     [(r+1) \Dscr_r]^\ell \Escr\, a_{r-1}^{jr + m +2} \, .
\cr
}}
\end{equation*}
Therefore, 
$$
\vcenter{\openup1\jot\halign{
 \hbox {\hfil $\displaystyle {#}$}
&\hbox {\hfil $\displaystyle {#}$\hfil}
&\hbox {$\displaystyle {#}$\hfil}\cr
\Fscr_{jr+m}^{(r)} &\le 
& \Escr \,a_{r-1}^{jr+m+2}\left[1+ [(r+1)\Dscr_r]^j  +
  \sum_{\ell=1}^{j-1}  \binom {j}{\ell}
  \frac {[(r+1)\Dscr_r]^\ell }{a_{r-1}^{r\ell}}\right]
\cr
&= & \Escr\, a_{r-1}^{jr+m+2} \sum_{\ell=0}^j \binom {j}{\ell}
\left(\frac{  (r+1)\Dscr_r}{a_{r-1}^r} \right)^{\ell}
\le 
\Escr\, a_{r-1}^{jr+m+2}\left( 1+ \frac{ (r+1)\Dscr_r}{a_{r-1}^r}\right) ^j 
\cr
&\le &
\Escr\, \left[a_{r-1}
  \left( 1+ \frac{ (r+1)\Dscr_r}{a_{r-1}^r}\right)^{\frac 1 r}\right]^{jr + m +2} \,.
\cr
}}
$$
This concludes the proof of the case $s>r$, not being a multiple of $r$.

Let us now consider $s= jr$ with $j\ge 2$. In such a case,
it is convenient to rewrite formula~\eqref{frm:nuovi-pert}
in the following more extended form:
$$
f_{jr}^{(r)} \hookleftarrow
\frac{1}{j!} \Lie {\chi_r}^{j} Z_0
+ \frac{1}{(j-1)!} \Lie {\chi_r}^{j-1} f_r^{(r-1)}
+\sum_{\ell=1}^{j-2} \frac{1}{\ell!}\Lie {\chi_r}^\ell f_{(j-\ell)r}^{(r-1)}\, ,
$$
where we have separated the term generated by $f_r^{(r-1)}$. Since
$\chi_r$ solves the homological equation~\eqref{frm:eq-omologica}, we
can rewrite the first two contributions as follows:
\begin{align*}
  \frac{1}{j!} \Lie {\chi_r}^{j} Z_0 +
  \frac{1}{(j-1)!} \Lie {\chi_r}^{j-1} f_r^{(r-1)} &= 
  \frac{1}{j!} \Lie {\chi_r}^{j-1} (\Lie {\chi_r} Z_0 + f_r^{(r-1)})
  + \frac{j-1}{j!} \Lie {\chi_r}^{j-1} f_r^{(r-1)} \\
  & = \frac{1}{j!} \Lie {\chi_r}^{j-1} Z_r
  + \frac{j-1}{j!} \Lie {\chi_r}^{j-1} f_r^{(r-1)} \, .
\end{align*}
Therefore, for $j \ge 2$ we can write
\begin{align*}
  \Fscr_{jr}^{(r)} & \le \Fscr_{jr}^{(r-1)} +
  \frac{1}{j} \left\| \frac{1}{(j-1)!} \Lie {\chi_r}^{j-1} Z_r \right\|
  + \frac{j-1}{j} \left\| \frac{1}{(j-1)!}
  \Lie {\chi_r}^{j-1} f_r^{(r-1)}\right\|
  + \sum_{\ell=1}^{j-2} \left\| \frac{1}{\ell!}
  \Lie {\chi_r}^\ell f_{(j-\ell)r}^{(r-1)} \right\|\\
  & \le \Fscr_{jr}^{(r-1)}+ \frac{1}{j} \cdot
  \frac{\prod_{i=0}^{j-2}(r + ir +2)}{(j-1)!}\Dscr_r^{j-1} \| Z_r \|\\
  & \phantom{\le}+ \frac{(j-1)}{j} \cdot
  \frac{\prod_{i=0}^{j-2}(r + ir +2)}{(j-1)!} \Dscr_r ^{j-1}
  \left\| f_r^{(r-1)} \right\| \\
  & \phantom{\le} + \sum_{\ell=1}^{j-2}
  \frac{\prod_{i=0}^{\ell-1}((j-\ell)r + ir +2)}{\ell!} \Dscr_r^\ell
  \left\| f_{(j-\ell)r}^{(r-1)} \right\| \\
  & \le \Fscr_{jr}^{(r-1)} + \frac{1}{(j-1)!} \prod_{i=0}^{j-2}[(1 + i)r +2)]
  \Dscr_r^{j-1} \Fscr_r^{(r-1)} \\
  & \phantom{\le} + \sum_{\ell=1}^{j-2} \frac{1}{\ell!}
  \prod_{i=0}^{\ell-1}[(j-\ell+i)r + 2]\Dscr_r^\ell \Fscr_{(j-\ell)r}^{(r-1)} \\
  & \le \sum_{\ell=0}^{j-1} \frac{1}{\ell!} \prod_{i=0}^{\ell-1}[(j-\ell+i)r + 2]
  \Dscr_r^\ell \Fscr_{(j-\ell)r}^{(r-1)} \, ,
\end{align*}
where we used again inequality~\eqref{frm:stima-lie} and
$\Fscr_r^{(r-1)}$ as upper bound for the norm of $Z_r\,$.  The second
inequality in formula~\eqref{frm:stima-per-code-r} follows from
calculations similar to those we have previously described for the
case where $s$ is not a multiple of $r$.
\end{proof}

\noindent
Let us emphasize that the same scheme of estimates is valid also for a
resonant Birkhoff normal form, since we did not use any explicit
definition of the terms $Z_s$ appearing
in~\eqref{frm:hr-expansion}. Therefore, in order to cover the
extension to such a case, it will be enough to adapt the definition of
the smallest divisor $\alpha_r$ and, as a consequence, the estimate
for the generating function $\chi_r\,$.

\section{Computer-assisted proofs about effective stability}
\label{sec:eff-stab}
Here we are going to use the scheme of estimates described in the
previous section for producing results on the effective stability in
the neighborhood of an elliptic equilibrium point. We will study
separately two different cases: a non-resonant Birkhoff normal form
and a resonant one. For what concerns the effective stability in the
 vicinity of equilibrium points, up to the best of our knowledge other
 computer-assisted rigorous estimates are available
 in~\cite{Cel-Gio-1991} only, but such a reference has to be considered
 rather outdated, because of some technical reasons affecting the
 final quality of the results (see also the comments in the
 Introduction). Moreover, an application of this kind of approach to a
 resonant Birkhoff normal form is entirely new.

\subsection{Estimates on the escape time for non-resonant Birkhoff normal forms}
\label{subsec:time-nonres}
Let us suppose that the Hamiltonian has been already brought in
Birkhoff normal form, after having performed $r$ normalization steps,
i.e., $\Hscr^{(r)} = \Zscr^{(r)} + \Rscr^{(r)}$ with $\Zscr^{(r)}$
depending on the actions only and
$\Rscr^{(r)}=\Oscr({\|\vet{I}\|}^{(r+3)/2})$. Therefore, we can
calculate the time derivative of the action $I_j =z_j \bar z_j$ as
follows:
\begin{equation}
\label{frm:derivata-azione}
\dot I_j=  \left\{I_j,\Hscr^{(r)}\right\}= 
\left\{I_j, \Zscr^{(r)}+ \Rscr^{(r)}\right\}=  
\left\{I_j, \Rscr^{(r)}\right\}
\quad \forall \ j=1, \ldots, n \, , 
\end{equation}
because $\{I_j\,,\,\Zscr^{(r)}(\vet{I})\}=0$. Using the
expansion~\eqref{frm:hr-expansion}, the estimate on the $\sup$ norm
in~\eqref{frm:stima-norma-sup} and lemma~\ref{lem:stima-poisson}, we
can find the following upper bound for the sup norm of the time
derivative of the actions\footnote{The general formula in
  lemma~\ref{lem:stima-poisson} would give a factor $2$, but here we
  take advantage of the fact that $I_j = z_j \bar z_j$ so we do not
  have terms with $z_j^2$. }:
\begin{equation}
\label{frm:stima_der}
\vcenter{\openup1\jot\halign{
 \hbox {\hfil $\displaystyle {#}$}
&\hbox {\hfil $\displaystyle {#}$\hfil}
&\hbox {$\displaystyle {#}$\hfil}\cr
|\dot I_j | _\rho &= &\sup_{\Delta_{\rho } }|\dot I_j |  =  \sup_{\Delta_{\rho } } \left| \left\{I_j,\sum_{s = r+1}^{+ \infty} f_s^{(r)}\right\} \right| \le 
\sum_{s=r+1}^{+ \infty}\, \rho^{s+2}  \|\poisson {I_j} {f_s^{(r)}} \|
\cr
& \le  & \sum_{s=r+1}^{+\infty} (s+2) \rho^{s+2} \| f_s^{(r)}\| \,.
\cr
}}
\end{equation}
We can now provide rigorous estimates for the norms of all the terms
appearing in the previous series. More precisely, we can use the
estimates $\Fscr_s^{(r)}$ (as recursively defined
in~\eqref{frm:stime-ridef}) for the first $R_{\rm II}$ terms.
Moreover, for what concerns the first $R_{\rm I}$ terms, we can do
even better, because we have explicitly performed $R_{\rm I}$ steps of
the normalization algorithm; therefore, for such terms the estimates
can be replaced by the actual values of the corresponding norms. For
all the remaining terms, we limit ourselves to use the uniform
estimate in~\eqref{frm:stima-per-code-r}. Thus, we just need to
compute the value of the upper bound in the r.h.s. of the following
inequality:
\begin{equation}
\label{frm:stima_der_calc}
\vcenter{\openup1\jot\halign{
 \hbox {\hfil $\displaystyle {#}$}
&\hbox {\hfil $\displaystyle {#}$\hfil}
&\hbox {$\displaystyle {#}$\hfil}\cr
\sum_{s=r+1}^{+\infty} (s+2) \rho^{s+2} \| f_s^{(r)}\| &\le & \sum_{s=r+1}^{R_{\rm I}}(s+2)\rho^{s+2} \|f_s^{(r)}\| + \sum_{s=R_{\rm I}+1}^{R_{\rm II}}(s+2)\rho^{s+2} \Fscr_s^{(r)} \cr
& & +\Escr \sum_{s= R_{\rm II}+1}^{+\infty} (s+2)\rho^{s+2} a_r^{s+2} \,.
\cr
}}
\end{equation}
Since the remainder is expected to be exponentially small at the
optimal step, it is clear that the quantity in~\eqref{frm:stima_der}
can be very small as well. From a computational point of view,
iterating estimates is much cheaper than doing algebraic calculations
on huge expansions; this is why such a procedure can be very
convenient when we are near the equilibrium point, where the optimal
normalization step is far beyond the number of steps we are able to
explicitly perform, by using any kind of software that is specialized
for doing computer algebra.

Let us now suppose that the initial value of the action vector belongs
to an open domain, e.g., $\vet I( 0)\in \Delta_{\rho_0}\,$; we aim to
determine a time $T>0$ (as long as possible) such that we can ensure
that $\vet I(t)\in \Delta_{\rho}$ $\forall\ t\in[-T\,,\,T]$, being the
domain $\Delta_{\rho}$ a little larger than $\Delta_{\rho_0}\,$,
i.e. $\rho>\rho_0\,$.  First, let us remark that
\begin{equation}
\label{frm:stima-nek}
 | I_j(t) | \le 
 | I_j(0) | + |I_j(t)- I_j(0)| \le 
\rho_0^2 +  |\dot I_j|_\rho \, T \, \quad  \forall \ j =1,\, \ldots\,,\, n.
\end{equation}
When an estimate for $|\dot I_j|_\rho \ \forall\ j$ is available, it
is convenient to define $T$ as
\begin{equation}
\label{frm:intermediate-time}
\frac{\rho^2 - \rho_0^2}{\max_j|\dot I_j|_\rho} 
\end{equation}
to obtain the wanted confinement. Combining~\eqref{frm:stima_der}
and~\eqref{frm:stima_der_calc} with~\eqref{frm:intermediate-time}, we
obtain the following lower bound about the escape time from
$\Delta_{\rho}\,$:
\begin{equation}
\label{frm:time}
T(\rho, \rho_0, r) =
\frac{\rho^2-\rho_0^2}
     {{\displaystyle{\sum_{s=r+1}^{R_{\rm I}}}}(s+2)\rho^{s+2} \|f_s^{(r)}\| +
      {\displaystyle{\sum_{s=R_{\rm I}+1}^{R_{\rm II}}}}(s+2)\rho^{s+2} \Fscr_s^{(r)} +
       \Escr {\displaystyle{\sum_{s= R_{\rm II}+1}^{+\infty}}}(s+2)\rho^{s+2} a_r^{s+2}}
     \, ,
\end{equation}
where the dependency on different parameters of such an expression is
emphasized; moreover, $r=r_{\rm opt}$ is chosen so as to minimize the
estimate~\eqref{frm:normasup-resto} of the remainder $\Rscr^{(r)}$. It is
rather easy to verify that the value 
of $\rho_0$ making a further optimization\footnote{When $r=r_{\rm opt}\,$, for the sake of
   simplicity, let us assume that the geometrical decrease of the terms
   in the series appearing at the denominator of
   equation~\eqref{frm:time} is so sharp that it can be approximated by
   its first term, then we can write $T(\rho, \rho_0, r_{\rm
     opt})\simeq g(\rho)$, being $g(\rho)=C(\rho^2-\rho_0^2)/[(r_{\rm
       opt}+3)\rho^{(r_{\rm opt}+3)}]$ and $C$ a suitable positive
   constant. After having remarked that
   $g(\rho_0)=\lim_{\rho\to\infty}g(\rho)=0$, one immediately realizes that
   the function $g:[\rho_0\,,\,\infty)\mapsto\reali$ takes its maximum
     value in correspondence to the solution of the equation
     $g'(\rho)=0$, i.e., $\rho = [{{(r_{\rm opt}+3)}/{(r_{\rm
             opt}+1)}}]^{1/2}\,\rho_0\,$.} of the expression above is
   $\rho_0 = [{{(r_{\rm opt}+1)}/{(r_{\rm opt}+3)}}]^{1/2}\,\rho$.

Therefore, in applications where the value of $\rho$ is considered as
fixed, we will compute the lower bound for the escape time
$T\big(\rho\,,\,[{{(r_{\rm opt}+1)}/{(r_{\rm
        opt}+3)}}]^{1/2}\,\rho\,,\,r_{\rm opt}\big)$ according to
formula~\eqref{frm:time}, where $r_{\rm opt}$ makes optimal the
estimate for the remainder $\Rscr^{(r)}$.

Let us recall that, in the framework of purely analytical estimates,
the denominator appearing in formula~\eqref{frm:time} would be made
just by its third summand with the series starting from $r+1$ instead
of $R_{\rm II}+1$; moreover, one can verify that $a_r\sim
(r!)^{\tau+1}C^r$, being $C$ a suitable positive constant. By choosing
$r=r_{\rm opt}$ as in formula~\eqref{frm:passo-ottimale}, one can
obtain the asymptotic law for the purely analytical estimate about the
escape time from $\Delta_{\rho}\,$, i.e.,
\begin{equation}
  \label{frm:stima_pura_analitica_tempo_stabilita}
  T\sim\exp\left[\left(\frac{\tilde\rho}{\rho}\right)^{\frac{1}{\tau+1}}\right]\ ,
\end{equation}
where $\tilde\rho$ is a positive constant. For more details we refer
to theorem~3.5 in~\cite{Giorgilli-2003.1}, where a complete proof of
such a lower bound can be found.

\subsection{Resonant Birkhoff normal forms: bounds on the escape time}
\label{subsec:time-res}

In the case of a resonant Birkhoff normal form, the estimate of the
diffusion of the actions is not so straightforward as for the
non-resonant one. Indeed, we cannot use directly the remainder in
order to estimate the variation of the resonant actions as we did
in~\eqref{frm:stima_der}. Here, we will discuss the case in which
there is only one resonant angle.

Let us assume we have already performed the construction of a resonant
Birkhoff normal form up to order $r$, in such a way that the
Hamiltonian of the system has the following structure:
\begin{equation}
\label{frm:forma-normale-ris}
\Hscr^{(r)} (\vet I, \vet \phi) =
\Zscr^{(r)}(\vet I,  \phi_n) +
\Rscr^{(r)}(\vet I, \vet \phi)\, ,
\end{equation}
where $\vet I \in \reali^{n}$ and $\vet \phi \in \toro^{n}$ are
action--angle canonical coordinates, being $\phi_n$ the resonant
angle.  Therefore, the estimates in~\eqref{frm:stima_der}
and~\eqref{frm:stima-nek} still hold true for what concerns the
diffusion of the actions $I_1\,,\, \ldots\,,\, I_{n-1}\,$, since we
removed the corresponding angles from the normal form. On the other
side, it is more difficult to provide good estimates for the diffusion
of the action $I_n\,$, that is conjugated to the resonant angle
$\phi_n\,$. Since the computation of the stability time $T$ as in
formula~\eqref{frm:time} is valid when all the actions are confined in
the domain $\Delta_\rho$, it is necessary to control the variation of
the action $I_n$ in order to use such an estimate about the value
of~$T$.  For this purpose, it is convenient to combine the
conservation of the total energy of the system, i.e., the Hamiltonian,
with the slow diffusion of the actions $I_1, \ldots, I_{n-1}\,$. More
precisely, by omitting the dependency on the index $r$ in
formula~\eqref{frm:forma-normale-ris}, we can write the following
equation:
\begin{equation}
\label{frm:In}
I_n = \frac{E - \sum_{j=1}^{n-1} \omega_jI_j}{\omega_n} - 
\frac{{\bar\Zscr}_1(\vet I,  \phi_n) + \Rscr(\vet I, \vet \phi)}{\omega_n}\,,
\end{equation}
where $E$ is the energy level and we denote with ${\bar\Zscr}_1$ the
sum of all the normal form terms but the linear ones with respect to
the actions (that are still of the form $\vet\omega\cdot\vet I$ as in
 the initial Hamiltonian~\eqref{frm:h0-iphi}). By assuming the confinement of all the actions in a ball of radius $\rho^2$, we can evaluate an upper bound of
\begin{equation}
\label{frm:max_In}
I_{n;\,{\rm max}} = \max_{|t|\le T}\,\sup_{\Delta_\rho}\left\{
\frac{E - \sum_{j=1}^{n-1} \omega_jI_j}{\omega_n} - 
\frac{{\bar\Zscr}_1(\vet I,  \phi_n) + \Rscr(\vet I, \vet \phi)}{\omega_n}
\right\}\,,
\end{equation}
and a lower bound of
\begin{equation}
\label{frm:min_In}
I_{n;\,{\rm min}} = \min_{|t|\le T}\,\inf_{\Delta_\rho}\left\{
\frac{E - \sum_{j=1}^{n-1} \omega_jI_j}{\omega_n} - 
\frac{{\bar\Zscr}_1(\vet I,  \phi_n) + \Rscr(\vet I, \vet \phi)}{\omega_n}
\right\}\,.
\end{equation}
This allows us to give an estimate of
\begin{equation}
  \label{frm:deltaIn}
  \Delta I_n = I_{n;\,{\rm max}} - I_{n;\,{\rm min}}\,.
\end{equation}
Of course, the excursions experienced by the values of the normal form
$\Zscr$ mainly depend on the linear terms; this explains why we have
written them separately in the equations above.  The bounds for the
sup-norm and inf-norm appearing in
formul{\ae}~\eqref{frm:max_In}--\eqref{frm:min_In}, respectively, can
be easily calculated, by using the
inequalities~\eqref{frm:stime-norme-termini}
and~\eqref{frm:stima-per-code}. Since the normal form is composed by a
finite number of terms, the estimate for $\Delta{\bar\Zscr}_1$ can be
evaluated very strictly when its expansion is explicitly known.

Recalling that the estimate for the variation of $I_n$ holds true only
if all the actions are confined in a ball of radius $\rho^2$, the
maximal radius on $I_n$ in order to validate such an estimate is
defined as
\begin{equation}
  \label{frm:gen_def_rhostar}
  ({\rho^*_n})^2 = \rho^2 -\Delta I_n\ .
\end{equation}
Therefore, if we define $T$ as in
formula~\eqref{frm:intermediate-time} for $j=1, \ldots, n-1$, we can
assure that $\vet I(t)\in \Delta_\rho \ \forall \ t \in[-T,\ T]$,
provided that $|I_j(0)|< \rho_0^2 \ \forall\ j=1,\ldots, n-1$ and
$|I_n(0)|< (\rho^*_n)^2$. Let us remark that the more $\rho_0$ is
close to $\rho$, the more the variation of $I_n$ is prescribed to be
small; at the same time, as it follows immediately from
formula~\eqref{frm:time}, the stability time decreases for $\rho_0$
going to $\rho$.  In order to balance these two effects, it is
convenient to define the parameter $\rho_0$ in a different way to what
has been done at the end of the previous subsection. In the next
section, we will discuss in a practical example how the radius
$\rho_0$ of the open ball containing the initial conditions can be
determined in the case of resonant Birkhoff normal forms.

\section{Effective stability of a couple of physical models}
\label{sec:results}

\subsection{An application to the H\'enon-Heiles model}
The H\'enon-Heiles model was introduced in~\cite{Hen-Hei-1964} for
studying the dynamics of a star in a galaxy. The Hamiltonian that
describes the system is composed of a couple of harmonic oscillators
perturbed with cubic terms, i.e.,
\begin{equation}
\label{frm:ham-HH}
\Hscr(\vet y,\vet x) = \omega_1 \frac{x_1^2 +y_1^2}{2}
+ \omega_2 \frac{x_2^2 +y_2^2}{2} + x_1^2x_2 -
\frac{x_2^3}{3}\qquad {\rm with} \ (\vet y, \vet x)\in \reali^{2n}\,.
\end{equation}
Here, we consider the non-resonant case only; in particular, we fix
$\omega_1=1$ and $\omega_2 = -(\sqrt 5-1)/2$, so that the angular
velocity vector is Diophantine with $\tau=1$ according to its
definition in~\eqref{frm:non-res}, because the ratio of its components
$|\omega_1/\omega_2|$ is equal to the golden mean\footnote{The
  prominent role exerted by the so called noble numbers is highlighted
  in~\cite{Mac-Sta-92}; the golden mean is the main representative of
  such a class of numbers, forming a subset of the Diophantine ones
  with $\tau=1$.}. We focus on the study of the dynamics in a
neighborhood of the origin.  This Hamiltonian belongs to the general
class described in~\eqref{frm:h0}, for which we have explained how to
perform the normalization procedure {\it \`a la} Birkhoff.

Using {{\it X}$\rho${\it \'o}$\nu${\it o}$\zeta$}, which is an
algebraic manipulator specially designed to implement approaches that
are common in the framework of Hamiltonian perturbation theory
(see~\cite{Gio-San-Chronos-2012}), we explicitly constructed the
Birkhoff normal form for this model by representing terms having total
polynomial degree less than or equal to $102$ and by performing
$R_{\rm I}=100$ normalization steps.  Using the iterative estimates
described in section~\ref{sec:stime}, combined with the norms that
have been determined for the first $R_{\rm I} = 100$ terms (for which
we have explicit expansions), we provided rigorous estimates for the
remainder and the time derivative of the actions, for any fixed value
of $\rho$.  Indeed, for the first $R _{\rm II}= 1500$ terms we
iterated the estimates of the norms as described in
proposition~\ref{pro:itera-stime-singole}, while for the rest of the
infinite terms we used the upper bounds reported in
formula~\eqref{frm:stima-per-code}, updating the value of $a_r$ as
prescribed in~\eqref{frm:a_r}; finally, we produced an estimate for
both the remainder and the escape time $T$, by using
formul{\ae}~\eqref{frm:normasup-resto} and~\eqref{frm:time}. For each
value of $\rho$ we considered, the computational algorithm was stopped,
after having identified the optimal step $r_{\rm opt}$ as the one
corresponding to the minimum value of the remainder; thus, we fixed
our final estimate for the escape time in such a way that $T=T(\rho,
[{{(r_{\rm opt}+1)}/{(r_{\rm opt}+3)}}]^{1/2}\,\rho, r_{\rm opt})$, in
agreement with the discussion at the end of
subsection~\ref{subsec:time-nonres}.

Let us recall that a similar technique, with the same meaning of the
integer parameters $R _{\rm I}$ and $R _{\rm II}\,$, allowed to obtain
a fully rigorous computer-assisted proof of existence for the KAM
torus related to the golden mean ratio of frequencies in the case of
the forced pendulum with a value of the small parameter that is
$~\sim\,92$\% of the breakdown threshold
(see~\cite{Cel-Gio-Loc-2000}); as far as we know, this is still the
best result of such a very particular kind, for what concerns the
applications of KAM theory to a Hamiltonian continuous flow.  In order
to fit with a so successful approach, in our programming codes we have
implemented validated numerics exactly in the same way. Therefore,
every coefficient appearing in all the polynomial expansions that are
explicitly computed (for Hamiltonians or generating functions) are
replaced with intervals and all the mathematical operations between
intervals are performed by taking into account the round-off
errors. For what concerns the iteration of the estimates, validated
numerics is used in order to properly provide all the needed
bounds. In summary, the whole computational procedure is aiming at
ensuring that the final result is fully
rigorous. Appendix~\ref{app:interval} is devoted to a short
introduction to validated numerics, that is described with the only
goal of explaining our implementation of it.  Furthermore, in
appendix~\ref{app:example} we have included a pedagogical discussion
of a computer-assisted estimate of the stability time for the
H\'enon-Heiles model in a very simple situation: we have focused on a
so small neighborhood of the equilibrium point (i.e., we have fixed
$\rho=0.0001$) that a few normalization steps are enough to obtain a
final result that is poor but still meaningful.

\input{tab.tex}

\begin{figure}
\centering
\subfigure{\includegraphics[width=7.9cm]{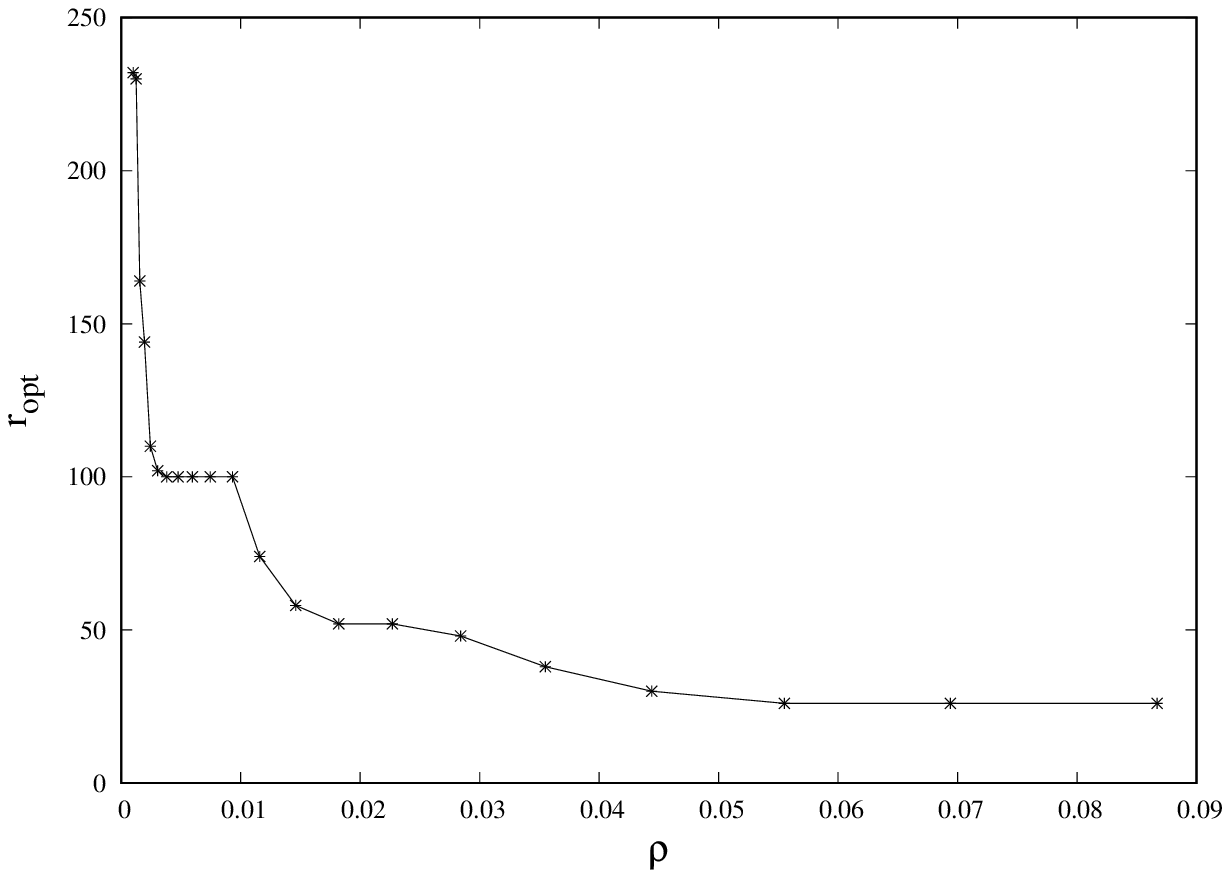}}
\subfigure{\includegraphics[width=7.9cm]{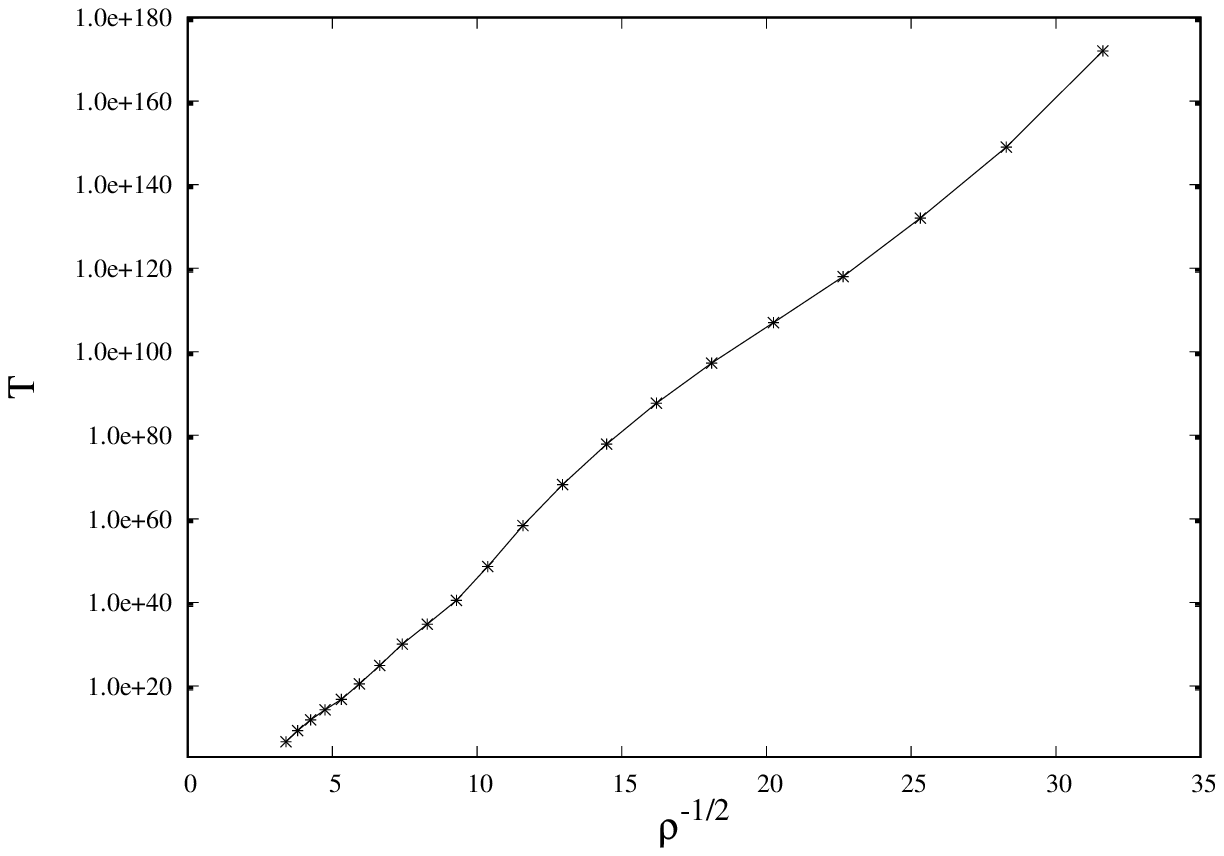}}
\caption{On the left, plot of the optimal normalization step
  $r_{\rm opt}$ as a function of the ball radius $\rho\,$; on the right,
  graph of the evaluation of our lower bound about the escape time $T$
  as a function of $1/\sqrt{\rho}\,$. Both the plots refer to results
  obtained by applying computer-assisted estimates to the
  H\'enon-Heiles model with frequencies $\omega_1 = 1$ and $\omega_2 = -
  (\sqrt 5 -1)/2$.}
\label{fig:rho_ropt_e_tempi}
\end{figure}

In Table~\ref{tab:chronos+stime} we have included our results for some
values of the distance $\rho$ from the equilibrium point: let us
emphasize that each row corresponds to a single computer-assisted
proof. Therefore, we think that every item of that list can be useful
for comparisons with the results eventually obtained by using other
techniques. Moreover, in Figure~\ref{fig:rho_ropt_e_tempi} we have
reported the main content of Table~\ref{tab:chronos+stime}, that is
about the behaviour of both the optimal step and the estimate for the
escape time as functions of $\rho$. In the left box of
Figure~\ref{fig:rho_ropt_e_tempi} one can appreciate that the limit
value of $\rho$ for which we are able to produce stability results is
around $0.1\,$. This limitation is due to the rate growth of $a_r\,$,
whose values get close to~$10$ just after a few normalization steps;
this prevents the convergence of the normal form, when $\rho\gtrsim
0.1$, because the common ratio of the majorants series is too large:
$a_{r}\rho \ge 1$. Looking at the definition in~\eqref{frm:a_r}, one
immediately realizes that having sharp estimates for the generating
functions is crucial to obtain good final results. This fact is
emphasized also by the occurrence of the plateau in the plot of the
function $r_{\rm opt}(\rho)$, that is in correspondence with the
number of normalization steps $R_{\rm I}=100$ for which we explicitly
compute the expansions. Indeed, the worsening effect induced by the
transition to the mere iteration of the estimate is so remarkable that
we have to strongly decrease the ball radius $\rho$ in order to take a
real advantage of performing more normalization steps. With the only
noticeable exception of the already mentioned plateau, the plot of the
optimal normalization step looks consistent with the expected law,
i.e., $r_{\rm opt}(\rho)\sim 1/\sqrt{\rho}$ (see
formula~\eqref{frm:passo-ottimale}). The agreement between the
observed behavior and the expectations is even better for what
concerns the plot in the right box of
Figure~\ref{fig:rho_ropt_e_tempi}. In fact, from the asympotic
law~\eqref{frm:stima_pura_analitica_tempo_stabilita} one can deduce
that $\log T\sim 1/\sqrt{\rho}$, that is coherent with the fact that
such a semi-logarithmic plot fits rather well with a straight
line. Thus, we can conclude that our computer-assisted estimates about
(the lower bound of) the stability time preserve its main property:
being exponentially large with respect to the inverse of the square
root of the distance from the equilibrium point.

Let us conclude this subsection with a short discussion about the
choice of the parameters $R_{\rm I}$ and $R_{\rm II}\,$. Obviously, it
is convenient to perform the largest possible number of explicit
steps. Therefore, our choice of the parameter $R_{\rm I}$ is mainly
due to the computational cost in time and memory that is needed by our
algebraic manipulations. The choice of the parameter $R_{\rm II}$ is
more delicate, because it somehow rules the size of the ``infinite
tail" of terms in the series. For small values of $\rho$, the main
contribution to the remainder~\eqref{frm:normasup-resto} is due to the
low order terms and the series of common ratio $a_r \rho$ does not
affect the result in an appreciable way, even for small values of
$R_{\rm II}\,$. In such a situation, choosing larger values of $R_{\rm
  II}$ has the only effect to increase the computational time, without
any substantial improvement for what concerns the estimate of the
remainder.  Conversely, for larger values of $\rho$, the common
ratio $a_{r}\rho$ can approach $1$ and the contribution of the
infinite tails becomes predominant. Therefore, it is convenient to
increase $R_{\rm II}$ in order to estimate more properly the
remainder.  This explains why, in our calculations, we have usually
fixed $R_{\rm II}=1500$, but we have suitably increased such a value
when $a_{r}\rho\lesssim 1$. However, the computational cost of the
iteration of the estimates is negligible; to fix the ideas, when
$R_{\rm II}=1500$ it took less than a minute, while performing
explicitly $R_{\rm I}=100$ normalization steps required about one day
of CPU-time on a computer equipped with an {\tt Intel quad-core
  I5-6600 - 3.3~Ghz} and {\tt 16~GB of RAM}. Let us recall that such
an impressive difference of computational cost is due to the fact that
the representation of the polynomials defined by the normalization
algorithm requires a huge occupation of the memory if compared with a
single upper bound on the corresponding norm and, therefore, many more
operations are needed to manipulate their explicit Taylor
expansions. For planning new applications it is also important to know
the scaling law of the computational time $T_{ \rm CPU}$ as a function
of $R_{\rm I}\,$: we have found that for the first $100$ normalization
steps such a law is not extremely sharp but it can be bounded from top
(and conveniently approximated) in such a way that $T_{\rm CPU}\sim
R_{\rm I}^8\,$.

\subsection{Applications to the asteroidal motions of the Trojans}
Let us briefly recall the Hamiltonian model we start from (that is
described, e.g., in~\cite{Giorgilli-1997}
and~\cite{Gab-Jor-Loc-2005}), in order to study the long term
stability in the vicinity of the triangular Lagrangian equilibria for
the CPRTBP.  As a first preliminary step, a rotating frame $Oxy$ with
origin $O$ in the centre of mass of the two primary bodies and rescaled
physical units is introduced. The Hamiltonian of such a system is
usually written as follows (see, e.g.,~\cite{Szebehely-1967}):
$$
  H(p_x,p_y,x,y) =
  \frac{p_x^2 + p_y^2}{2} + y p_x -x p_y- \frac{1-\mu}{\sqrt{(x-\mu)^2+y^2}}
  -\frac{\mu}{\sqrt{(x+1-\mu)^2+y^2}}\ ,
$$
where $\mu$ is the mass ratio between the primaries and
$(p_x\,,\,p_y)\in \reali^2$ are the kinetic momenta that are
canonically conjugate to the positions $(x,y)\in \reali^2$,
respectively.  Therefore, it is convenient to perform a preliminary
transformation to heliocentric polar coordinates; afterwards, we
introduce local coordinates $(P_X,P_Y,X,Y)$ centered in a triangular
point. They are defined in such a way that
$$
\vcenter{\openup1\jot\halign{
 \hbox {\hfil $\displaystyle {#}$}
&\hbox {\hfil $\displaystyle {#}$\hfil}
&\hbox {$\displaystyle {#}$\hfil}
\ \ &\hbox {\hfil $\displaystyle {#}$}
&\hbox {\hfil $\displaystyle {#}$\hfil}
&\hbox {$\displaystyle {#}$\hfil}\cr
  x &=& (X+1) \cos\Big(Y\mp\frac{2\pi}{3}\Big)+\mu\>,
  & p_x &=& P_{X} \cos\Big(Y\mp\frac{2\pi}{3}\Big)
  - \frac{\sin\big(Y\mp\frac{2\pi}{3}\big)}{X+1}(P_Y+1)\>,
  \cr
  y &=& (X+1) \sin\Big(Y\mp\frac{2\pi}{3}\Big)\>,
  & p_y &=& P_{X} \sin\Big(Y\mp\frac{2\pi}{3}\Big)
  + \frac{\cos\big(Y\mp\frac{2\pi}{3}\big)}{X+1}(P_Y+1)+\mu\>,
  \cr
}}
$$
where minus and plus refer to the equilibrium points $L_4$ and
$L_5\,$, respectively.  In these new variables, the Hamiltonian takes
the form
$$
\vcenter{\openup1\jot\halign{
 \hbox {\hfil $\displaystyle {#}$}
&\hbox {\hfil $\displaystyle {#}$\hfil}
&\hbox {$\displaystyle {#}$\hfil}\cr
H(P_X,P_Y,X,Y) &=& \frac{1}{2} \left[ P_X^2+\frac{(P_Y+1)^2}{(X+1)^2}\right]
-P_Y -\mu (X+1)\cos\Big(Y\mp\frac{2\pi}{3}\Big) - \frac{1-\mu}{X+1}
\cr
& & -\frac{\mu}{\sqrt{(X+1)^2+1+2(X+1)\cos\Big(Y\mp\frac{2\pi}{3}\Big)}}\ .
\cr
}}
$$
After having skipped a constant term, the basic Taylor expansion of
the previous Hamiltonian can be written as $H(P_X,P_Y,X,Y) =
\Qscr(P_X,P_Y,X,Y)+\Oscr\big(\big|(P_X,P_Y,X,Y)\|^3\big)$, where the
terms that are quadratic with respect to the canonical variables are
gathered in $\Qscr$.  It is well known that if $\mu$ is smaller than
the so called Routh-Gascheau critical value $(9-\sqrt{69})/18$, then
the equation of motion are linearly stable. This means that there
is a class of linear canonical transformations\footnote{A procedure
  determining such a canonical transformation $\Cscr$ can be found in
  section~7 of~\cite{Gio-Del-Gal-Sim-1989}. The code allowing
  Mathematica to compute the symplectic matrix related to $\Cscr$ is
  freely available at the web address {\tt
    http://www.mat.uniroma2.it/$\sim$locatell/MCSH/programmi/diag\_L4oL5.mth}}
$(P_X,P_Y,X,Y)=\Cscr(y_1, y_2,x_1,x_2)$ conjugating the quadratic
approximation to a couple of harmonic oscillators, i.e.,
$\Qscr\big(\Cscr(y_1,y_2,x_1,x_2)\big)=
{\nu_1}(x_1^2+y_1^2)/{2}+{\nu_2}(x_2^2+y_2^2)/{2}$. Therefore, the
Taylor series expansion of the Hamiltonian $\Hscr=H\circ\Cscr$ can be
written in the same form described in~\eqref{frm:h0}, that is suitable
to start the normalization procedure {\it \`a la} Birkhoff, i.e.,
\begin{equation}
  \Hscr(y_1, y_2,x_1,x_2) =
       {\nu_1}  \frac {x_1^2+y_1^2}{2}+ {\nu_2} \frac {x_2^2+y_2^2}{2} +
       \sum_{\ell = 3}^{+\infty} f_\ell(y_1,y_2,x_1,x_2)
       \quad {\rm with}\ f_\ell \in \Pscr_\ell\,,
\end{equation}
being the angular velocities $\nu_1$ and $\nu_2$ such that
\begin{equation}
\label{frm:frequenze}
\nu_1 = \sqrt{\frac {1+\sqrt{27\mu^2-27\mu+1}}{2}}\, \quad {\rm and}\quad
\nu_2 =-\sqrt{\frac {1-\sqrt{27\mu^2-27\mu+1}}{2}} \ .
\end{equation}
As an immediate consequence, we have that $\nu_1\to 1$ and
$\nu_2=\Oscr(\mu)$ for $\mu\to 0$.  The second frequency is therefore
much slower than the first one, for small values of the mass ratio
$\mu$ between the primaries. Let us remark that during the standard
procedure constructing the Birkhoff normal form (that has been widely
discussed in section~\ref{sec:birk}), very small divisors can be
introduced because of the fact that $|\nu_2|\ll 1$. Therefore, it is
natural to expect that a resonant Birkhoff normal form, aiming to
remove just the first angle (i.e., the fastest) can be more
advantageous. Let us also recall that the bodies orbiting around a
triangular equilibrium point are commonly called ``Trojans'',
according to the tradition started by Max Wolf at the beginning of the
XX~century. In fact, he decided to choose names from Homer's Iliad
for the first bodies which were observed in the vicinity of $L_4$ or
$L_5\,$, in the system having Sun and Jupiter as primary bodies.

In the next subsection we will compare the performances of these two kinds
of Birkhoff normal forms, by considering a few realistic values for
the parameter $\mu$, which correspond to systems having the following
couples of primary bodies: Sun--Jupiter, Sun--Uranus, Sun--Mars and
Saturn--Janus. In particular, we aim to prove the stability for a time
that is comparable with an overestimate of the residual life of the
Sun in the main sequence, i.e., $6\times10^9$ years.

\subsubsection{Effective stability of trojan celestial bodies in the Solar system}
\begin{figure}
\centering
\subfigure{\includegraphics[width=7.9cm]{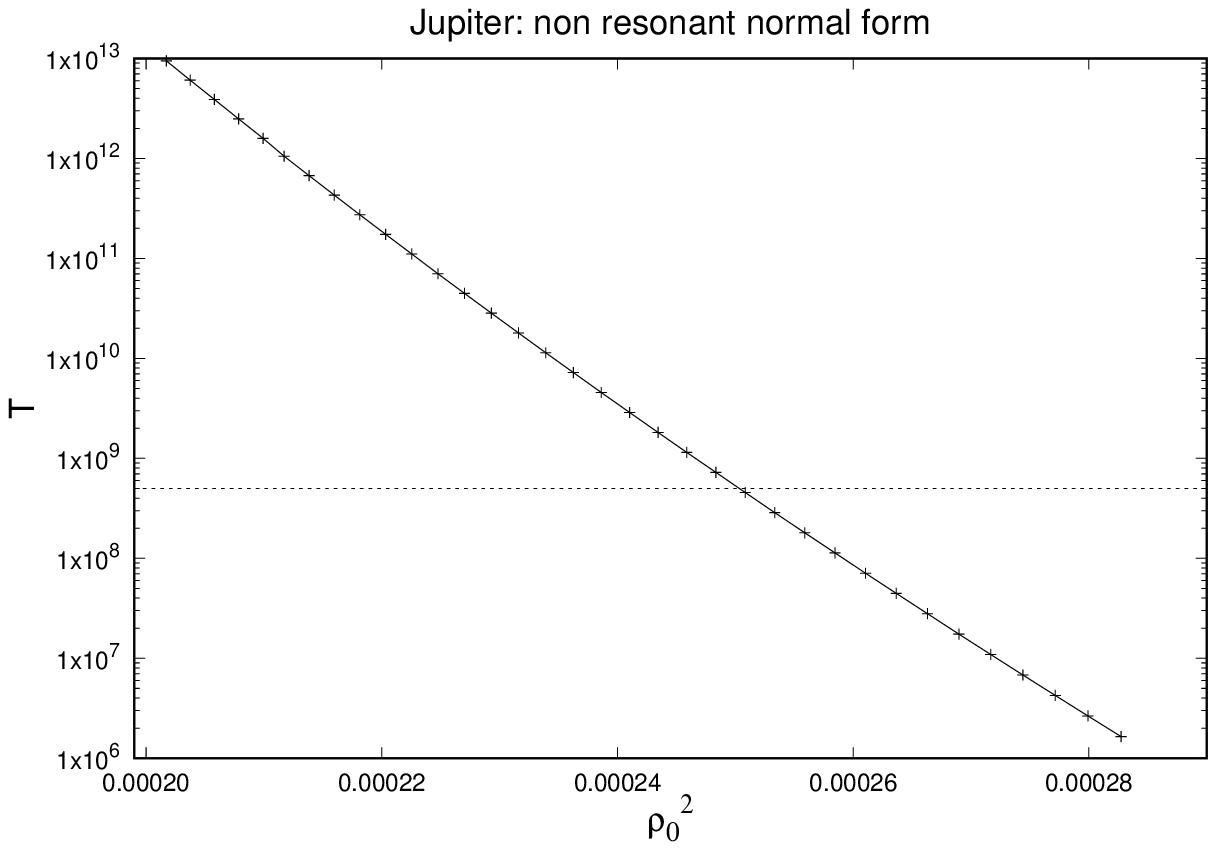}}
\subfigure{\includegraphics[width=7.9cm]{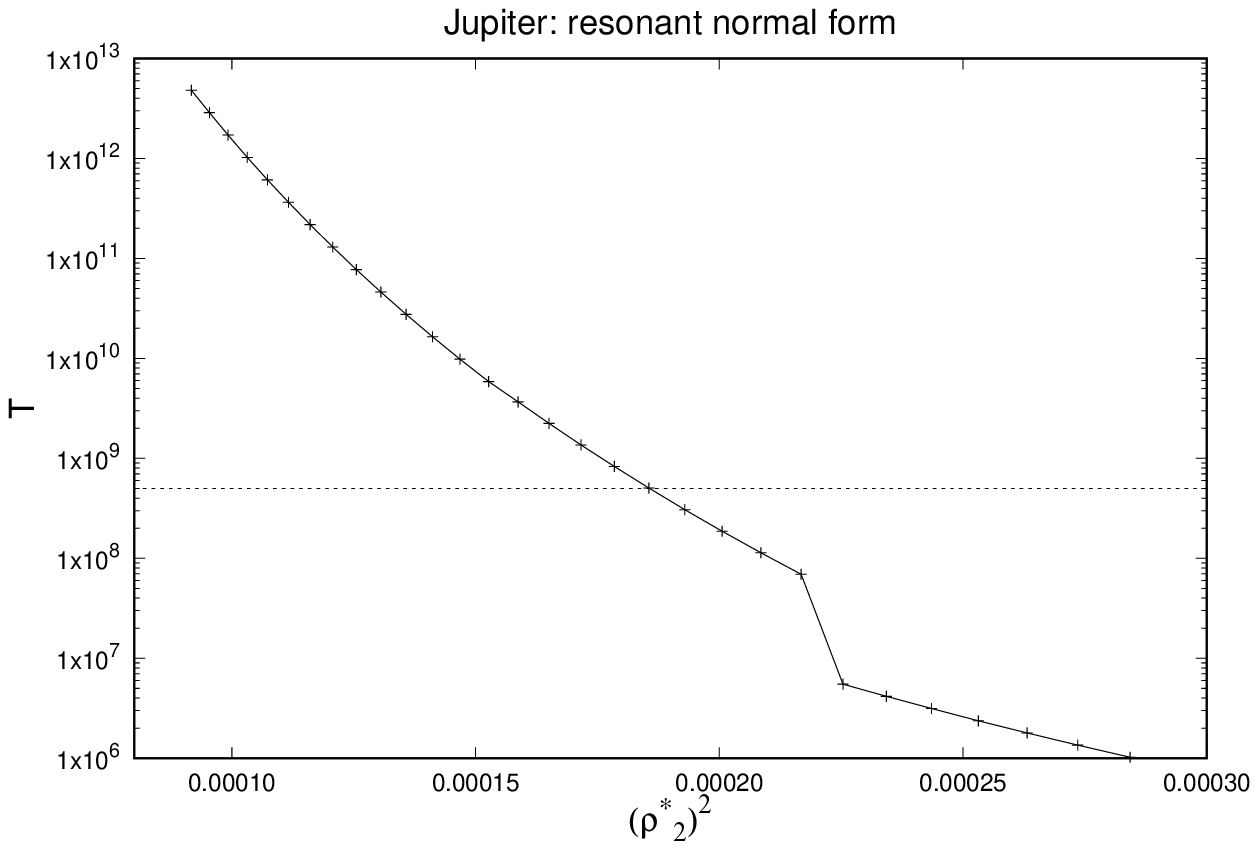}}
\caption{Plots of the evaluation of our lower bound of the escape time
  $T$ (in semi-log scale). On the left, the graph is a function of
  $\rho_0\,$, on the right, of ${\rho^*_2}\,$. The horizontal line
  corresponds to $T_{ \rm e. l. t.} = 5\times 10^8$. See the text for
  more details.}
\label{fig:rho_tempi}
\end{figure}

As it has been discussed in the introduction, an approach merely based
on the Birkhoff normal form is not enough to obtain realistic results
about the stability of the Jupiter Trojans in the framework of the
CPRTBP model. In~\cite{Giorgilli-1997}, it has been shown that orbital
motions starting from initial conditions contained in the domain
$(I_1,I_2)\in[0,0.0008]\times[0,0.0005]$ are effectively stable. This
is far from explaining why the regions in the proximity of $L_4$ and
$L_5$ are so populated; indeed, that domain covers the observational
data of a very small fraction of Trojans. The results get even worse
when all our rigorous estimates about both the remainder and the
stability time are taken into account.  As it is clearly shown in the
left hand side of Table~\ref{tab:jupcomp}, the construction of the
non-resonant Birkhoff normal form allows us to ensure the effective
stability for initial conditions such that
$(I_1,I_2)\in[0,0.00025]\times[0,0.00025]$.  In that table we reported
only a couple of results for two different (and very close) values of
$\rho$.  In fact, for this particular problem, it has no physical
meaning to prove stability for a time longer than the life-time of the
Sun. Therefore, as it has been done in~\cite{Giorgilli-1997}, we
examine the values of $\rho$ for which the stability time is
comparable with the so called expected life-time $T_{ \rm e. l. t.} $
of the system. This number is rescaled with respect to the revolution
period of the celestial body, i.e., we impose that $T_{ \rm e. l. t.}
= (6\ {\rm Gyrs})\cdot \frac{\nu_1}{2\pi}\simeq 5\times 10^8$ (in
number of Jupiter revolutions, because $\nu_1$ is expressed in {\rm
  rad$/$yrs}). As it has been highlighted in
Figure~\ref{fig:rho_tempi}, a small decrease of the value of the radii
containing the initial condition $\rho_0$ can very significantly
change the stability time (because of the exponential dependency of
$T$ on the inverse of $\rho$).  Such a behaviour is evident also in
Table~\ref{tab:jupcomp}, where with a change of the $1\%$ of $\rho$,
the stability time is almost doubled in the non--resonant case.  Let
us emphasize that in our estimates, we use the same radius for both
the actions, while looking at the initial conditions of the real
trojan asteroids for Jupiter in~\cite{Gab-Jor-Loc-2005}, the value of
the second action is usually smaller with respect to the first
one. Therefore, we believe that considering the initial conditions of
the actions in polydisks of different radii for the actions, as
in~\cite{Giorgilli-1997}, could improve the results.

\input{Jupcomp.tex}

In order to make easier the comparisons, in our opinion it is
convenient to report the results about both the non-resonant normal
form and the resonant one next each other, as we have done in
Table~\ref{tab:jupcomp} and also in Figure~\ref{fig:rho_tempi}. For
the sake of definiteness, we have to explain how to determine the
radius $\rho_0$ of the open ball containing the initial
conditions. Let us recall that at the end of
subsection~\ref{subsec:time-nonres} we have chosen a value of $\rho_0$
in order to optimize our evaluation of the lower bound of the escaping
time $T$. Instead, here we focus our attention on the consistency of
the procedure, in the sense that we aim to ensure that the motion is
confined in the domain $\Delta_\rho$ for all $|t|\le T$.  Therefore,
we fix $\rho_0\in(0\,,\,\rho)$ as the unique solution of the equation
\begin{equation}
  \beta \rho^2 =
  \rho^2 + \frac{ \nu_1 (\rho^2 - \rho_0^2)}{ \nu_2}\ ,
\end{equation}
where $\nu_1$ and $\nu_2$ have opposite signs in view
of~\eqref{frm:frequenze} and $\beta <1$ is another parameter we have
the freedom to choose, because it essentially represents the price we
are willing to pay in terms of a further restriction concerning the
domain of the initial value of the resonant action $I_2(0)\,$.  This
last concept definitely deserves a more detailed explanation.  Indeed,
in agreement with~\eqref{frm:gen_def_rhostar}, we define
$({\rho^*_2})^2=\rho^2-\Delta I_2\,$, with $\Delta I_2$ given as
in~\eqref{frm:deltaIn}.  Since the main contribution to the maximal
variation of the resonant action is due to the linear terms (see
formul{\ae}~\eqref{frm:max_In}--\eqref{frm:deltaIn}), we have that
$\Delta I_2 \simeq - {\nu_1 (\rho^2 - \rho_0^2)}/{ \nu_2}$ and,
therefore, $({\rho^*_2})^2 \simeq \beta\rho^2$.  Let us rephrase a
conclusion discussed at the end of section~\ref{subsec:time-res}, by
adapting it to the present context: we have that $\vet I(t)\in
\Delta_\rho \ \forall \ t \in[-T,\ T]$, provided that $|I_1(0)|<
\rho_0^2$ and $|I_2(0)|<(\rho^*_2)^2\simeq\beta\rho^2$. This ends the
explanation on the meaning of $\beta$ as the parameter ruling the
restriction on the set of values of $I_2(0)$. For what concerns the
choice of the value of $\beta$, we are interested in
setting it very close to $1$, in order to enlarge the domain of
stability in $I_2(0)$ as much as possible; on the other side, we have
to take into account oscillations of the value of the resonant action
that are due to the angular dependence of the normal form part
${\bar\Zscr}_1(\vet{I},\phi_2)$ (see
formul{\ae}~\eqref{frm:max_In}--\eqref{frm:min_In}). Indeed, this so
called modulating term for the resonant action $I_2$ is such that
${\bar\Zscr}_1=\Oscr(\|\vet{I}\|^2)$. This remark allows to imagine
a further optimization procedure on the determination of $\beta$, that
in principle would not be so difficult, due to the already mentioned
fact that the maximal variation $\Delta I_2$ of the resonant action is
linear with respect ot the the actions $(I_1\,,\,I_2)$ in view of
formul{\ae}~\eqref{frm:max_In}--\eqref{frm:deltaIn}.  For the sake of
simplicity, we prefer to avoid such a further optimization procedure:
in all the systems we have studied in the framework of the CPRTBP
model, we have simply set $\beta = 0.9\,$. Indeed, we consider that an
eventual further enlargement of the domain of stability in $I_2(0)$
is not so crucial, because it cannot be greater than 10~\%; therefore,
it cannot substantially improve our results, as it is clearly shown in
the summary of the comparisons reported in the final
Table~\ref{tab:birk-res-birk}.

\begin{figure}[h]
\centering
\subfigure{\includegraphics[width=7.9cm]{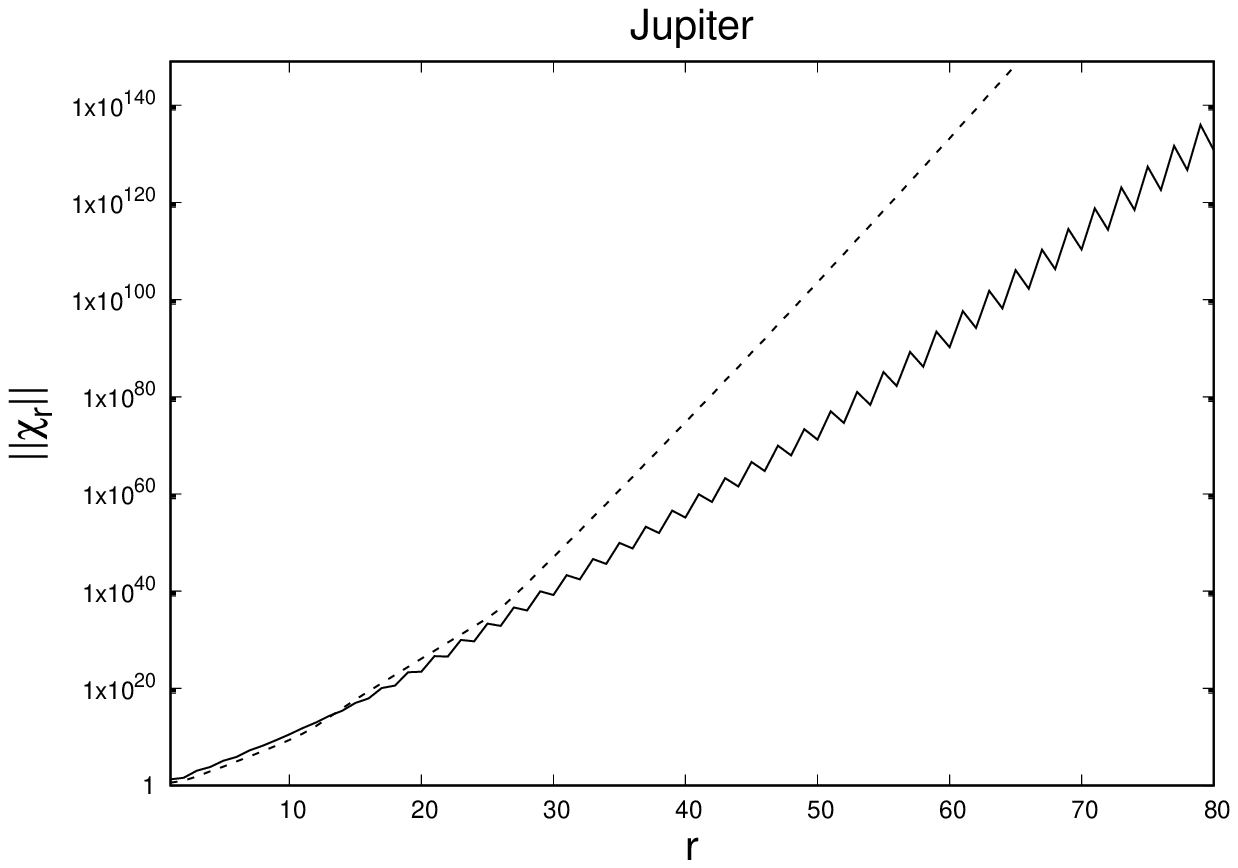}}
\subfigure{\includegraphics[width=7.9cm]{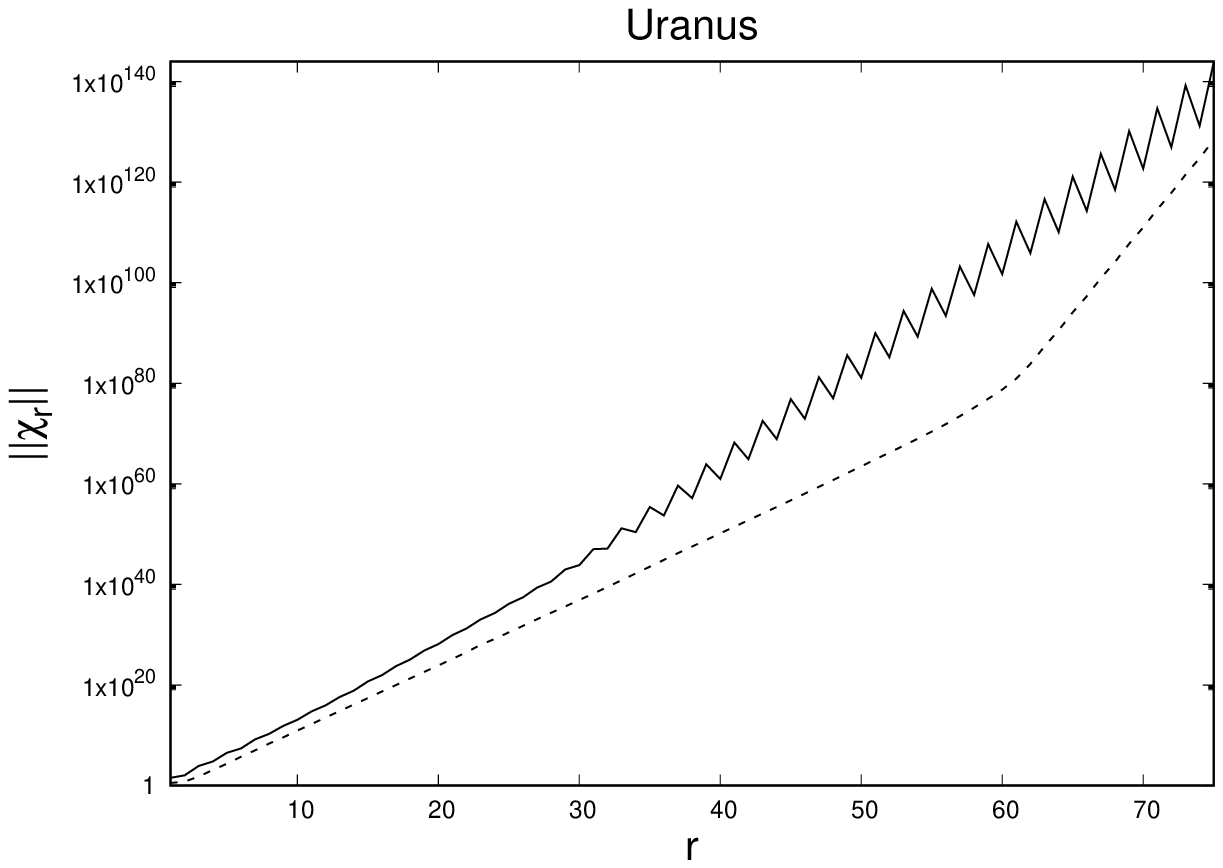}}
\subfigure{\includegraphics[width=7.9cm]{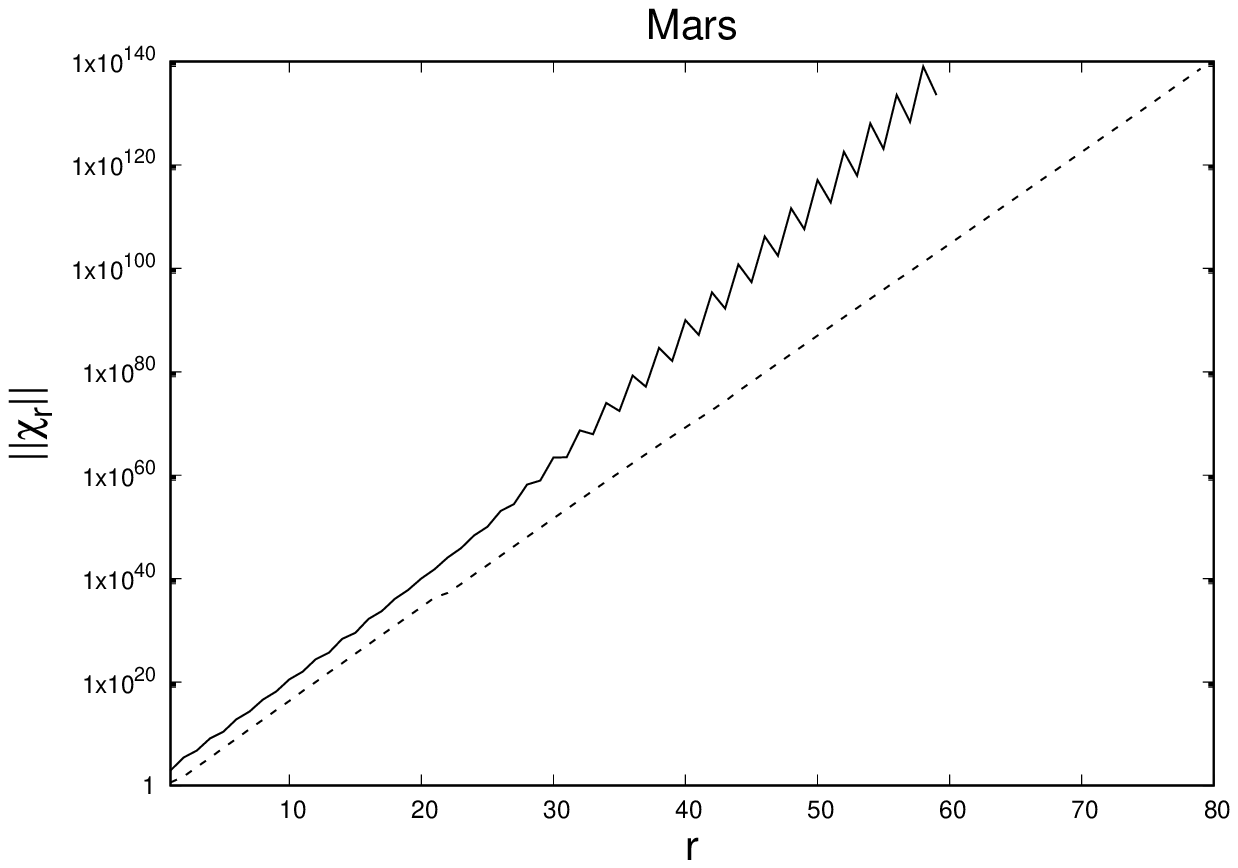}}
\subfigure{\includegraphics[width=7.9cm]{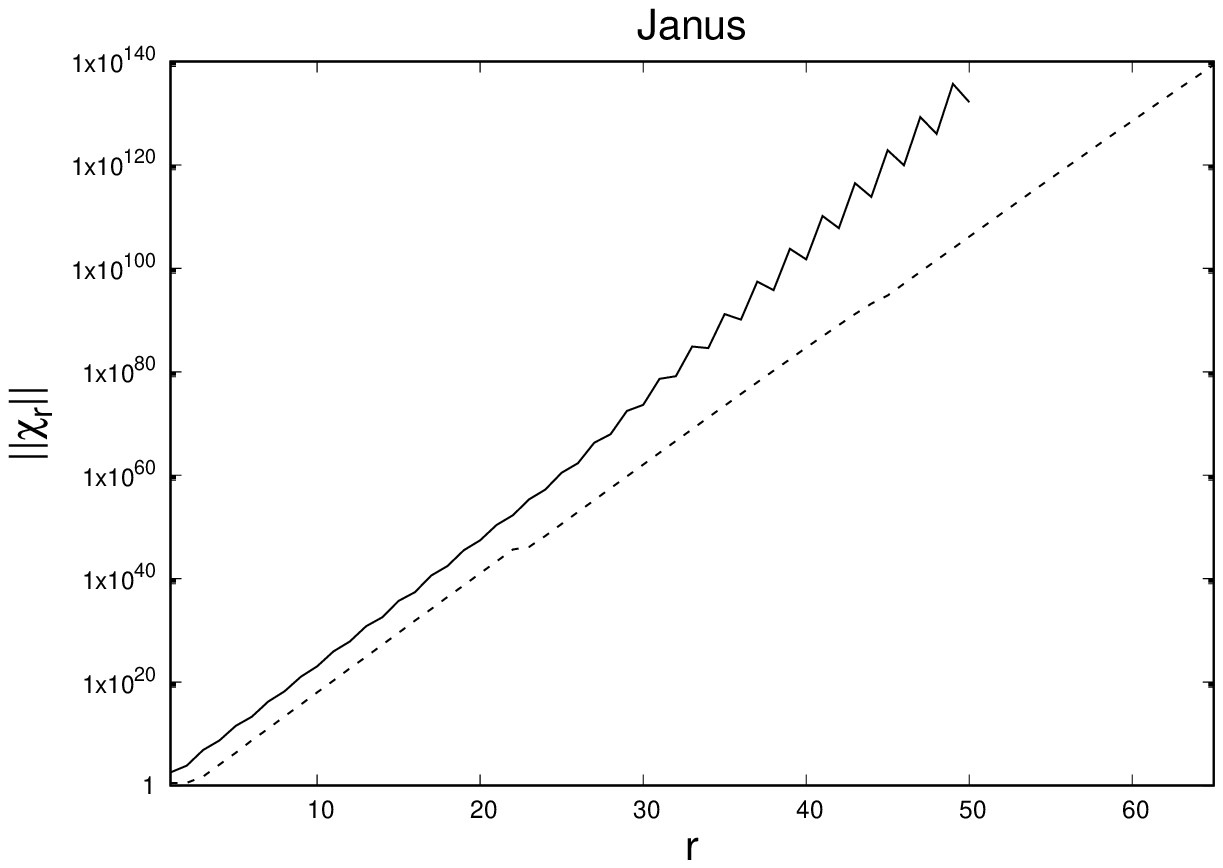}}
\caption{Growth of the norms (in semi-log scale) of the generating
  functions for the non-resonant Birkhoff normal form (continuous
  line) and the resonant one (dashed line). From top to down and from
  left to right, the boxes refer to the cases of the systems having
  Sun--Jupiter, Sun--Uranus, Sun--Mars and Saturn--Janus
  as primary bodies, respectively.}
\label{fig:confgen}
\end{figure}

The right hand side of Table~\ref{tab:jupcomp} includes the results
based on the construction of a resonant Birkhoff normal form, whose
performance is worse with respect to the non-resonant one. The
difference of behaviour between the two methods can be explained, by
looking at the top-left box in Figure~\ref{fig:confgen}, where there
is the comparison between norms of the generating functions $\chi_r$
for the resonant and non-resonant constructions, in the case of the
system having Sun and Jupiter as primary bodies.
The increase of the norms proceeds more and less at the same rate up
to the first $25$ steps; afterwards, the norms of the generating
functions related to the resonant normal form start to grow up faster
than the other ones. On the one hand, while the algorithm for the resonant
normal form is performed, there are less terms where small divisors
are introduced by the solution of the homological
equation~\eqref{frm:eq-omologica}; on the other hand, the normal form
part is larger in the case of the resonant construction and this
clearly has an impact on the size of the generating function. In fact,
when the $r$--th normalization step is performed, new perturbative
terms are introduced by the application of Lie series to the
Hamiltonian $\Hscr^{(r-1)}$; the main ones are of type $\Lie
{\chi_r}Z_k$ for $k<r$. Let us recall that in the non--resonant
construction we have $Z_i=0$ for all odd values of index $i$;
therefore, in that case we have a lower number of new terms  and the
main new contribution is expected to be due to $\Lie {\chi_r}Z_2\,$.
This explains why in the non-resonant case the growth of the
generating functions has two different trends for odd and even
indices, as it is clearly shown by the plots reported in
Figure~\ref{fig:confgen}. On the contrary, in the resonant case the
behaviour of the norms of the generating functions is more regular, at
least up to a threshold value of $r$, for which the worsening effect
due to the terms generated by the normal form part becomes predominant
with respect to the advantage gained because of the better
accumulation of small divisors.

\input{Uracomp.tex}
\input{Marcomp.tex}
\input{Jancomp.tex}

Let us now focus on the comparison between the performances of the
resonant and non-resonant constructions, when they are made for
different values of $\mu$: the results are summarized in the
top-right, bottom-left and bottom-right boxes of
Figure~\ref{fig:confgen} and in
Tables~\ref{tab:Uranus}--\ref{tab:Janus}, which refer to the systems
having Sun--Mars, Sun--Uranus and Saturn--Janus as primary bodies,
respectively. These systems have been chosen because of two reasons:
all of them host at least one trojan body and the corresponding values
of the mass ratio $\mu$ allow a study where such a parameter spans
rather regularly several orders of magnitude. In all the plots of
Figure~\ref{fig:confgen} (except the one referring to the case of the
Jupiter Trojans) the size of the generating functions produced by the
resonant construction algorithm is remarkably smaller with respect to
the corresponding non-resonant ones. Therefore, the series introduced
by the resonant normalization procedure are convergent in a bigger
neighbourhood of the origin; in other words, in such a situation we
are able to iterate the computer-assisted estimates for larger values
of $\rho$.

The results described in the present subsection are further summarized
in Table~\ref{tab:birk-res-birk}, where the records are listed in
decreasing order with respect to the mass ratio $\mu$.  This allows us
to emphasize that the comparison between the computer-assisted
estimates based on the non-resonant Birkhoff normal form and the
resonant one are more and more in favour of the latter, when the value
of $\mu$ tends to zero.
\begin{table}[t]
\begin{center}
  \begin{tabular}{c|c|c|c|c}
    & $\mu$ & $\rho_0^2\ \,{\rm (non-res.)}$
    & $({\rho^*_2})^2\ \,{\rm (reson.)}$
    & $({\rho^*_2}/\rho_0)^2$\\
\hline
Jupiter &$9.54 \times 10^{-4}$ &$2.49\times10^{-4}$ &  $1.83\times10^{-4}$ & 0.73\\
Uranus  &$4.36 \times 10^{-5}$& $8.30\times 10^{-5}$ & $7.57\times 10^{-4}$ & 9.12\\
 Mars & $3.21\times 10^{-7}$ & $7.36\times 10^{-6}$ &  $1.08\times 10^{-4}$& 14.67 \\
Janus &$3.36\times 10^{-9}$ & $6.00\times 10^{-7}$ &$1.10\times 10^{-5}$& 18.33 \\
\hline
\end{tabular}
\end{center}
\caption[]{Comparisons between the values of the radii $\rho_0^2$ and
  $({\rho^*_2})^2$ which refer to the stability domains for the
  non-resonant Birkhoff normal form and the resonant one,
  respectively. The results are reported as a function of different
  values of the mass ratio $\mu$, the name of the smaller primary in
  the corresponding CPRTBP model is reported in the first column.}
\label{tab:birk-res-birk}
\end{table}

\section{Conclusions and perspectives}
\label{sec:conclu}
Since the very beginning of our research project, one of our main
goals was to complement the Birkhoff normal form with a coherent
scheme of computer--assisted estimates, in order to provide rigorous
evaluations of the effective stability time. This has been
accomplished. Moreover, in our opinion we have developed a technique
that could be adapted in such a way to apply also in proximity of
equilibrium points that are not purely elliptical (see,
e.g.,~\cite{Jor-Mas-99} and~\cite{Joh-Tuc-09} to find examples of
interesting problems that could be studied with such an approach).

We have successfully applied our procedure to two simple systems,
which are close to elliptic equilibrium points such that the quadratic
approximation of the Hamiltonian is not convex (so preventing a
trivial proof of stability using the energy as Lyapunov function): the
H\'enon-Heiles model and the Circular Planar Restricted Three-Body
Problem (CPRTBP). For the latter, we have easily adapted our approach
to the construction of the Birkhoff normal form of resonant type and
we have shown that the consequent computer-assisted estimates are in a
better position when the mass ratio between the primary bodies is very
small.  Nevertheless, the extent of the domain covered by our results
is still far from being enough to explain the effective stability of
most of the Trojans.

In the near future, we plan to work for improving our
computer-assisted results, for what concerns the applicability to
realistic physical models, that are of interest, in particular, in
Celestial Mechanics and Astronomy. We think that in these fields there
still are problems that are open and fundamental\footnote{For
  instance, see C. Caracciolo, U. Locatelli, M. Sansottera, M. Volpi:
  ``Librational KAM tori in the secular dynamics of the
  $\upsilon$--Andromedae planetary system'', in preparation. A
  preliminary version is available on request to the authors.}, where rigorous proofs of stability can be tackled by using an
approach purely based on the normalization algorithm for KAM tori,
which can ensure a perpetual topological confinement in models with
two degrees of freedom. Nevertheless, there is a much wider range of
possible applications to Hamiltonian systems defined on phase spaces
of higher dimensions. In such a framework, a very promising strategy
is based on the local construction of the Birkhoff normal form in the
neighborhood of an invariant KAM torus: it has already started to
provide some remarkable results (see,
e.g.,~\cite{Gio-Loc-San-2017}). Here, in the context of the discussion
about the perspectives of our computational method, it is natural to
describe the scaling properties of our approach with respect to the
number~$n$, i.e., the degrees of freedom of the Hamiltonian model. It
is well known that the remainder of the Birkhoff normal form is
exponentially small with respect to the distance $\rho$ form the
equilibrium point; this allows to deduce the following best possible
estimate about the stability time:
$T\sim\exp[({\tilde\rho}/{\rho})^{{1}/{n}}]$, that is corresponding to
the most non-resonant frequencies in the small oscillations limit,
being $\tilde\rho$ a positive constant (recall
formul{\ae}~\eqref{frm:non-res}
and~\eqref{frm:stima_pura_analitica_tempo_stabilita}). Such an
asymptotic law is intrinsic to an approach based on Birkhoff normal
forms and we have shown that it is in agreement with the results
produced by our computational method. This is due to the fact that for
small distances $\rho$ the so called optimal normalization step
$r_{\rm opt}$ can largely exceed the maximal polynomial degree
$R_{\rm I}+2$ of the Hamiltonian terms whose expansions are explicitly
computed; in such a condition, our method starts to iterate the
estimates, i.e., a procedure that is very weakly affected by the
eventual increase of the number of degrees of freedom~$n$. Nonetheless, the accuracy of the final estimates about the stability
time remarkably deteriorates when $r_{\rm opt}$ becomes greater than
$R_{\rm I}$ (see Figure~\ref{fig:rho_ropt_e_tempi}). Increasing~$n$
strongly reduces the maximal integer value, say $R_{\rm I;max}$ that can
be conveniently fixed for the parameter $R_{\rm I}$ ruling the size of
the expansions stored on a computer. Since a polynomial function of
maximal degree $R_{\rm I}+2$ and depending on $2n$ variables hosts a
number of terms that is $\Oscr\big(R_{\rm I}^{2n}\big)$, it is natural
to expect the following scaling law for the threshold corresponding to
the deterioration of our computational results:
$R_{\rm I;max}\sim\Mscr^{1/(2n)}$, being $\Mscr$ a suitable positive
constant.

In the very particular case of the PCRTBP, there are other possible
sources of further improvements. Indeed, the normal form construction
that has been designed in~\cite{Pae-Loc-2015} adopted (since the very
beginning) canonical coordinates which are much more suitable for such
a kind of Celestial Mechanics model. This is the reason why the
approximation provided by that normal form describes much more
carefully the dynamics, when it is compared with the numerical results
based on the Birkhoff normal form. We think that our computer-assisted
estimates can be adapted to complement also the constructive algorithm
explained in~\cite{Pae-Loc-2015}. This should provide results about
the effective stability holding true in wider domains, hopefully covering
an important fraction of the trojan asteroids.

\subsection*{Acknowledgments}
This work was partially supported by the ``Mission Sustainability''
programme of the Universit\`a degli Studi di Roma ``Tor Vergata''
through the project IDEAS (E81I18000060005) and by the ``Progetto
Giovani 2019'' programme of the National Group of Mathematical Physics
(GNFM--INdAM) through the project ``Low-dimensional Invariant Tori in
FPU--like Lattices via Normal Forms''.  The authors acknowledge the
MIUR Excellence Department Project awarded to the Department of
Mathematics of the University of Rome ``Tor Vergata'' (CUP
E83C18000100006), in particular, because of the availability of the
computational resources.

\appendix
\section{Basics of validated numerics on a computer}
\label{app:interval}
In order to give a completely rigorous support to our
computer-assisted proofs, in the present work we have implemented
everywhere validated numerics in two complementary ways. First,
interval arithmetic has been used to calculate the coefficients of the
Taylor expansions for the normal forms (as far as possible), while
rigorous computations of upper [lower] bounds have been performed to
estimate other quantities of interest, e.g., norms of functions
[stability times, respectively].  The present appendix is devoted to
summarize our approach to validated numerics, that essentially follows
what is described in section~3 of~\cite{Koc-Sch-Wit-96}.

All our codes are written in {\bf C} language and all the quantities
of interest are computed using the {\tt double} type in {\bf C}.
The set of {\it floating point} numbers of {\tt double} type that are
representable on a computer is defined as follows:
\begin{equation}
  \label{frm:representable_numbers}
\vcenter{\openup1\jot\halign{
 \hbox {\hfil $\displaystyle {#}$}
&\hbox {\hfil $\displaystyle {#}$\hfil}
&\hbox {$\displaystyle {#}$\hfil}\cr
  \Rgot &= &\big\{0\big\}\,\cup\,
  \left\{\pm \left(\sum_{j=0}^{51}\frac{d_j}{2^{j}}\right)\cdot 2^{-1023}
  \,:\ d_j=0,\,1\>\forall\>j=0,\,\ldots\,51\right\}
  \cr
  & &\cup\,
  \left\{\pm \left(1+\sum_{j=1}^{52}\frac{d_j}{2^{j}}\right)\cdot 2^\sigma
  \,:\ d_j=0,\,1\>\forall\>j=1,\,\ldots\,52\,,\ \sigma\in\interi,
  \>-1022\le\sigma\le 1023\right\} \ ,
  \cr
}}
\end{equation}
where $\sigma$ is the {\it exponent} and the digits
$d_0.d_1d_2\,\ldots\,d_{52}$ make up the so called {\it
  mantissa}\footnote{In the first row of
  formula~\eqref{frm:representable_numbers}, $d_{52}$ is fixed to be
  zero, while $d_0$ is set to $1$ for the numbers appearing in the
  second row, according to the so called ``hidden bit'' rule. By
  taking into account that the integer exponents between $-1023$ and
  $1023$ require 11~{\it bits} to be represented and 1~{\it bit} is
  needed for the sign, it should be evident that the information to be
  stored for each {\tt double} type number must be spread over
  8~{\it bytes} (=64~{\it bits}).}  $m$; they appear in the binary
scientific notation $x=\pm m\,2^\sigma$ of each number $x$ belonging
to the set $\Rgot$. Therefore, an {\it overflow} [{\it underflow}]
situation can occur when a computational operation attempts to
generate a number whose absolute value is greater [smaller] than
$\Mgot=(1 - 2^{-53})\cdot 2^{1024}$ [than $\mgot=2^{-1074}$ and it is
different from $0$, respectively].
Let us generically use the symbol $\ast$ to refer to any elementary
arithmetic operations $+$, $-$, $\cdot$ and~$/$; moreover, the result
provided by a computer when it performs that same operation will be
denoted with $\circledast$. According to such a setting, for instance,
$a\oplus b$ is the computer output for the usual sum $a+b$, where both
$a$ and $b$ belong to $\Rgot$. Therefore, the {\it machine epsilon} for
{\tt double} type floating point numbers in {\bf C} (hereafter, simply
$\epsilon$) is defined as
$$
\epsilon=\min\left\{\eta\in\Rgot\,:\ 1\oplus\eta>1\right\}\ .
$$
By comparing the definition above with that
in~\eqref{frm:representable_numbers}, one can easily realize that
$\epsilon=2^{-52}$ and
$\epsilon/2=\min\big\{\eta\in\Rgot\,:\ 1\ominus\eta<1\big\}$.

The 64 bit IEEE standard ensures that, if $a\ast b\notin\Rgot$ and
$\mgot<|a\ast b|<\Mgot$, then either $a\circledast b =
\max\big\{\Rgot\cap(-\infty,\,a\ast b)\}$ or $a\circledast b =
\min\big\{\Rgot\cap(a\ast b,\,\infty)\}$; furthermore, if $a\ast
b\in\Rgot$, then $a\circledast b = a\ast b$. In words, these
prescriptions can be summarized as follows: when a generic elementary
operation does not create a situation of either {\it overflow} or {\it
  underflow}, then the result provided by a computer must be correct
up to the last significant digit\footnote{Also the square root
  $\sqrt{\phantom{1}}$ enjoys such a peculiar property (see, e.g.,
  section~3.1 of~\cite{Sch-Weh-Wit-2000}).\label{nota:radice_quadrata}}.

In order to prevent the occurrence of {\it overflows} or {\it
  underflows} generated by an elementary arithmetic operation
$\circledast$, we restrict to the ``safe range'' (according to
definition~3.2 in~\cite{Koc-Sch-Wit-96}), which is the following set
of numbers:
\begin{equation}
  \label{frm:safe_numbers}
  \Sgot = \big\{0\big\}\,\cup\,
  \left\{x\in\Rgot\,:\ 2^{-511}\le|x|\le 2^{511}\right\}\ .
\end{equation}
We can rephrase the prescriptions of the 64 bit IEEE standard for the
arithmetic elementary operations in a very quantitative way, by
referring to the relative error, i.e.,
$$
\forall\ a,b\in\Sgot\,:\ \ {\rm if}\ a\circledast b = 0\,,
\ \ {\rm then}\ a\ast b=0\,,
\ {\rm else}\ \left|\frac{a\ast b}{a\circledast b}-1\right|\le \epsilon\ .
$$
There is an obvious exception to such a general rule: the division
by zero. Of course, validated numerics is performed on a very restricted
domain with respect to the real numbers and it is not expected to extend
operations that are meaningless in $\reali$. Therefore, the computational
algorithm must be set in such a way that divisions by zero (and square
roots of negative numbers, etc.) are avoided.

Looking at the way the mantissa is represented in the second row of
formula~\eqref{frm:representable_numbers}, one can immediately realize
that for that type of floating point numbers multiplying by the factor
$1+\epsilon$ [by $1-\epsilon$] is enough to increase [decrease, resp.]
their absolute value by the last significant digit. This last
remark joined with the prescriptions of the 64 bit IEEE standard allows us to
establish a few simple rules, in order to rigorously perform interval
arithmetic. Let us introduce the following binary operators
$\circledast_+:\,\Sgot\times\Sgot\mapsto\Rgot$ and
$\circledast_-:\,\Sgot\times\Sgot\mapsto\Rgot$, that are defined as
follows:
\begin{equation}
  \label{def:aritmetica_elementare_con_arrotondamenti}
a\circledast_+ b =
\left\{
\vcenter{\openup1\jot\halign{
 \hbox {\hfil $\displaystyle {#}$}
 &\quad\hbox {$\displaystyle {#}$\hfil}\cr
 (a\circledast b)\odot(1+\epsilon)
 &{\rm if}\ a\circledast b\ge 0 \ ,
 \cr
 (a\circledast b)\odot(1-\epsilon)
 &{\rm if}\ a\circledast b < 0 \ ;
 \cr
}}
\right .
\qquad
a\circledast_- b =
\left\{
\vcenter{\openup1\jot\halign{
 \hbox {\hfil $\displaystyle {#}$}
 &\quad\hbox {$\displaystyle {#}$\hfil}\cr
 (a\circledast b)\odot(1-\epsilon)
 &{\rm if}\ a\circledast b\ge 0 \ ,
 \cr
 (a\circledast b)\odot(1+\epsilon)
 &{\rm if}\ a\circledast b < 0 \ .
 \cr
}}
\right .
\end{equation}
In view of the prescriptions of the 64 bit IEEE standard, for what
concerns the true results of any arithmetic elementary operation
we can conclude that
\begin{equation}
  \label{frm:fondamenti-di-aritmetica-rigorosa}
a\circledast_- b \le a\ast b \le a\circledast_+ b
\qquad
\forall\ a,b\in\Sgot\ ,
\end{equation}
where, again, $\ast$ generically denote $+$, $-$, $\cdot$ and $/$
(with the only exception of the division by zero). We emphasize that
the inequalities in
formula~\eqref{frm:fondamenti-di-aritmetica-rigorosa} are fundamental,
because they allow us to implement the rigorous computations of upper
bounds (or lower ones) for quantities that have to be
estimated. Moreover, the validated numerics can be applied to interval
arithmetic by suitably extending the binary operators introduced
in~\eqref{def:aritmetica_elementare_con_arrotondamenti}.  In fact, let
$\circledast_{\pm}:\,\Sgot^2\times\Sgot^2\mapsto\Rgot^2$ be defined in
such a way that $[a_{-}\,,\,a_{+}] \circledast_{\pm} [b_{-}\,,\,b_{+}]
= [c_{-}\,,\,c_{+}]$, where
\begin{equation}
  \label{def:limiti-rigorosi-intervalli}
\vcenter{\openup1\jot\halign{
 \hbox {\hfil $\displaystyle {#}$}
&\hbox {\hfil $\displaystyle {#}$\hfil}
&\hbox {$\displaystyle {#}$\hfil}\cr
  c_{-} &= &\min\big\{a_{-}\circledast_{-}b_{-}\,,\,a_{-}\circledast_{-}b_{+}\,,\,
                     a_{+}\circledast_{-}b_{-}\,,\,a_{+}\circledast_{-}b_{+}\big\}
  \cr
  c_{+} &= &\max\big\{a_{-}\circledast_{+}b_{-}\,,\,a_{-}\circledast_{+}b_{+}\,,\,
                     a_{+}\circledast_{+}b_{-}\,,\,a_{+}\circledast_{+}b_{+}\big\}
  \cr
}}
\ \>.
\end{equation}
Because of the definition of $\circledast_{\pm}$ and in view of
formul{\ae}~\eqref{frm:fondamenti-di-aritmetica-rigorosa}--\eqref{def:limiti-rigorosi-intervalli},
we can write the following relation that is fundamental for the (rigorous)
interval arithmetic:
\begin{equation}
  \label{frm:fondamenti-di-aritmetica-degli intervalli}
a\ast b \in [a_{-}\,,\,a_{+}] \circledast_{\pm}[b_{-}\,,\,b_{+}]
\qquad
\forall\ a\in[a_{-}\,,\,a_{+}]\,,
\ b\in[b_{-}\,,\,b_{+}]\ ,
\end{equation}
where, once again, the case with $0\in[b_{-}\,,\,b_{+}]$ and $\ast$
representing the division must be excluded.  Let us stress that two
(possibly redundant) list of elements appear after the minimum and the
maximum in the definitions~\eqref{def:limiti-rigorosi-intervalli} of
$c_{-}$ and $c_{+}\,$, respectively, in order to take into account of
the effects induced by the signs, when we are dealing with products
and divisions. To fix the ideas, it is convenient to realize that, for
instance, the rigorous extension of the sum to intervals, can be
defined in a much shorter way, i.e., $[a_{-}\,,\,a_{+}] \oplus_{\pm}
[b_{-}\,,\,b_{+}] = [a_{-}\oplus_{-}b_{-}\,,\,a_{+}\oplus_{+}b_{+}]$.
Since the algorithm constructing the Birkhoff normal form is based on
Lie series (therefore, Poisson brackets), one can immediately realize
that the computations of the coefficients appearing in the expansions
involve elementary arithmetic operations, only. Thus, their rigorous
extension to intervals is enough to perform validated numerics on the
truncated expansions of the Hamiltonian as far as they can be
explicitly computed. Let us stress that the whole computational
algorithm can be iterated provided that each generic arithmetic
operation of type $[a_{-}\,,\,a_{+}] \circledast_{\pm}
[b_{-}\,,\,b_{+}] = [c_{-}\,,\,c_{+}]$ generates a new result such that
$c_{-}\,,\,c_{+}\,\in\Sgot$. This is the reason why, in our codes, we
included tests, in order to verify that such a condition is always
satisfied; if it is not so, the running of a program is immediately
stopped.

The situation concerning the rigorous bounds on the estimates is
slightly more complicated, because, very quickly, it occurs a
violation of the condition that all the majorants of the norms are
given by numbers belonging to the ``safe range'' set $\Sgot$.  This is
because the estimates can be iterated for many more normalization
steps with respect to the explicit calculation of the expansions, each
of them requiring a huge occupation of the memory if compared with a
single upper bound on the corresponding norm. Due to the dramatically
fast growth of the norms of terms composing the series (that are
asymptotically diverging for the number of normalization steps going
to infinity), the limit $\max\Sgot$ is usually trespassed even for
values of $R_{II}$ that are relatively small with respect to those
considered in the applications described in
section~\ref{sec:results}. Let us recall that we are interested in
iterating the estimates of the norms for a large number of
normalization steps $R_{II}\,$, in order to obtain better results.
Therefore, it is convenient to represent the logarithm of the
(positive values of the) upper bounds of the norms. For such a
purpose, we have to introduce a function
$\log_+:\,\Sgot\cap\reali_+\mapsto\Sgot$ such that $\log x\lesssim
\log_+ x$ for all positive $x\in\Sgot$.  This can be done in a rather
obvious way, by adapting the approach described in section~3.1
of~\cite{Sch-Weh-Wit-2000}, in order to combine validated numerics
with a truncated series expansion of a logarithm. To fix the ideas,
let us limit to the case $x\in\Sgot\cap[1,\infty)$, then we
  define\footnote{Let us stress that in order to ensure that
  $\log x\le\log_{+}x$ where
  $\log_{+}x=2^n\odot_{+}\log_+\xi_n$ and provided that
  $\log\xi_n\le\log_{+}\xi_n$
  $\forall\ \xi_n\in\Sgot\cap[1\,,\,1.01]$, we exploit the fact
  that integer powers of two belong to $\Sgot$.}
$\log_{+}x=2^n\odot_{+}\log_+\xi_n\,$, being
$(\xi_n\ominus_+1)\in\Sgot\cap[0\,,\,0.01]$ such that
$\xi_0=x$ and $\xi_j=\big(\sqrt{\xi_{j-1}}\big)_+$
$\forall\ j=1,\ldots,n$, where $n$ is the minimum nonnegative integer such
that $\xi_n\le 1.01$ and the function $\big(\sqrt{\cdot}\big)_+$ is nothing
but $(1+\epsilon)\sqrt{\cdot}$, in agreement with what has
been explained in the corresponding
footnote${{\ref{nota:radice_quadrata}}\atop{\phantom{1}}}$.
The rigorous upper bound for
$\log\big(1+(\xi_n\ominus_+1))$ $\forall\ \xi_n\in\Sgot\cap[1\,,\,1.01]$
is introduced in a similar way to the (simpler) definition of the
function $\exp_{+}(\cdot)$ (see formula~\eqref{frm:def_exp_+} below).
The definition of $\log_{+}x$ for $x\in\Sgot\cap(0,1)$ is analogous to
the case discussed just above with $x\ge 1$.

The translation of the iterative estimates in terms of upper bounds on
the logarithms of the norms is obvious, when just products and
divisions are involved. For instance, let us focus on the last two
definitions appearing in the statement of
Proposition~\ref{pro:itera-stime-singole}, i.e.,
$\Dscr_r=(r+2)\Gscr_r\,$, with
$\Gscr_r={\Fscr_r^{(r-1)}}/{\alpha_r}\,$. In our codes we
can write the corresponding rigorous estimate as
$\log\Dscr_r=\log_{+}(r+2)\oplus_{+}\log{\Fscr_r^{(r-1)}}\oplus_{+}
\log_{+}\big({1}\oslash_{+}{\alpha_r}\big)\,$, where we mean that
$\log{\Fscr_r^{(r-1)}}$ is a previously defined number belonging to
the ``safe range'' set $\Sgot$. When algebraic sums are involved, the
procedure is slightly more complicate. As a further example, let us
focus on~\eqref{frm:a_r}: in order to properly define an upper bound
for the logarithm of $a_r=\big({a_{r-1}^r}+(r+1)\Dscr_r\big)^{1/r}$ we
have to compute $\log_{+}(x+y)$, where the values of
$\log x=r\odot_{+}\log a_{r-1}$ and
$\log y=\log_{+}(r+1)\oplus_{+}\log\Dscr_r$ have to be considered as
known, because both $\log a_{r-1}$ and $\log\Dscr_r$ have been
preliminarily estimated by some rigorous computations; moreover, let us
recall that we want to avoid to compute $x=\exp(\log x)$ and
$y=\exp(\log y)$, because they are expected to be too large numbers,
eventually exceeding $\max\Sgot$. Without any loss of generality, let
us assume that $x\ge y$, therefore, it is convenient to set
$$
\log_{+}(x+y)=\log x\oplus_{+}
\log_{+}\Big(1\oplus_{+}\exp_{+}\big(\log y\ominus_+\log x\big)\Big)\ .
$$
Thus, we are lead to the problem of defining a function
$\exp_+:\,\Sgot\cap[0,1]\mapsto\Sgot$ such that $\exp(x)\lesssim
\exp_+(x)$ $\forall\>x\in\Sgot\cap[0,1]$. Once again, we
follow~\cite{Sch-Weh-Wit-2000}; first, we introduce
$\xi_j=x\oslash_{+}2^j$ $\forall\ j=0,\ldots,n$, where $n$ is the
minimum positive integer such that $\xi_n\in\Sgot\cap[0\,,\,0.03]$.
Hence, we compute
\begin{equation}
  \label{frm:def_exp_+}
\exp_+(\xi_n) =
\left(\build{\scriptstyle{\bigoplus_{+}}}_{i=0}^{\Nscr}
({\pi_i}\oslash_{+}{i!})\right)\oplus_{+}
\Big(\big({\pi_\Nscr}\oslash_{+}{\Nscr!}\big)\oslash_{+}(1\ominus_{-}\xi_n)\Big)\ ,
\end{equation}
where $\build{\scriptstyle{\bigoplus_{+}}}_{i=0}^{\Nscr}({\pi_i}\oslash_{+}{i!})=
1\oplus_{+}\ldots\oplus_{+}({\pi_\Nscr}\oslash_{+}{\Nscr!})$ with
$\pi_i=\pi_{i-1}\odot_{+}\xi_n$ $\forall\ i=1,\ldots,\Nscr$ (being
$\pi_0=1$); moreover, $\Nscr$ is the minimum positive integer such that
${\pi_\Nscr}\oslash_{+}{\Nscr!}\le\epsilon$, while $\sum_{i>\Nscr}\xi_n^i/i!
\le\big({\pi_\Nscr}\oslash_{+}{\Nscr!}\big)\oslash_{+}(1\ominus_{-}\xi_n)$
is nothing but the estimate of the truncated remainder for the
expansion in Taylor series of $\exp(\xi_n)$. The rigorous computation 
of $\exp_+(x)=\exp_+(\xi_0)$ is completed by recursively defining
$\exp_+(\xi_j)=\exp_+(\xi_{j+1})\odot_{+}\exp_+(\xi_{j+1})$ in such
a way to proceed backwards from $j=\Nscr-1$ to $j=0$.

The whole of the elementary
operations\footnote{\label{nota:lib_val_num.c} These functions are
  included in the {\tt lib\_val\_num.c} library, that makes part of
  the package freely available at the web address {\tt
    http://www.mat.uniroma2.it/$\sim$locatell/CAPs/BirkCAP\_AppsA-B.zip}}
described in the present appendix represent the bare minimum that is
sufficient to implement validated numerics, in order to make fully
rigorous our computer-assisted proofs.

\section{A complete example of application to the H\'enon-Heiles
  model in a case with short expansions}
\label{app:example}
The aim of this appendix is very pedagogical: we explain step-by-step
an application of the algorithm described in the
sections~\ref{sec:birk}--\ref{sec:eff-stab}. For the sake of clarity,
here we focus again on the H\'enon--Heiles model (that is the simplest
one among those considered in section~\ref{sec:results}), by
performing a very low number of normalization steps. In fact, the
example described in the present appendix deals with the
case where $R_{\rm I}=2$ and $R_{\rm II}=5$; this means that the expansions
of larger polynomial degree to be explicitly computed are quartic
while the iteration of the estimates provides upper bounds for terms up to
the seventh degree. In our opinion this choice is a good balance: on the one hand it is non-trivial, on the other hand the representation of
the Hamiltonians is still readable because it is not too large.
Moreover, we have decided to further simplify our example, by omitting
the study of our better estimate for the stability time as function of
the radius $\rho$, whose value is fixed to $\rho=0.0001$.

First, let us emphasize that, during our computer-assisted proofs, a
Hamiltonian $\Hscr^{(r-1)}$ (the Taylor expansion of which is
explicitly given in equation~\eqref{frm:hr-1}) is represented by a
set\footnote{Let us stress that in formula~\eqref{frm:array} some
  unpleasant lower indexes (that are involving either the minimum or
  the maximum between nonnegative integers) are somehow unavoidable,
  because the number of already performed normalization steps,
  i.e. $r-1$, can be greater than $R_{\rm I}$ or not; therefore,
  either the explicit expansions of $f_{\min\{r,R_{\rm I}\}}^{(r-1)}
  \,,\,\ldots\,,f_{R_{\rm I}}^{(r-1)}$ are missing or the majorants
  $\log\Zscr_{R_{\rm I}+1} \,,\,\ldots\,,
  \log\Zscr_{\min\{r-1\,,\,R_{\rm II}\}}$ disappear in the list of
  elements making part of the set $\Sscr^{(r-1)}$.} gathering both
polynomial functions and numbers:
\begin{equation}
\label{frm:array}
\vcenter{\openup1\jot\halign{
 \hbox {\hfil $\displaystyle {#}$}
&\hbox {\hfil $\displaystyle {#}$\hfil}
&\hbox {$\displaystyle {#}$\hfil}\cr
 \Sscr^{(r-1)} &= \bigg\{
 &Z_0 \,,\,\ldots\,,  Z_{\min\{r-1,R_{\rm I}\}} \,,\,
 f_{\min\{r,R_{\rm I}\}}^{(r-1)} \,,\,\ldots\,,f_{R_{\rm I}}^{(r-1)} \,,
 \cr
 & &\log\Zscr_{R_{\rm I}+1} \,,\,\ldots\,, \log\Zscr_{\min\{r-1\,,\,R_{\rm II}\}} \,,\,
 \log\Fscr_{\max\{r,R_{\rm I}+1\}}^{(r-1)}\,,\,\ldots\,, \log\Fscr_{R_{\rm II}}^{(r-1)}\,,
 \cr
 & & \log\Escr \,,\, \log a_{r-1} \bigg\}
 \ .
 \cr
}}
\end{equation}
The polynomials appearing in the first row of definition~\eqref{frm:array}
are written in terms of their expansions, that are
$Z_{s}=\sum_{2|\vet \ell| = s}
c_{\vet \ell,{\vet \ell};\pm}^{(s)}\,
(-i \vet z)^{\vet \ell}\bar {\vet z}^{{\vet \ell}}$
and
$f_{s}^{(r-1)}=\sum_{|\vet \ell|+ |\tilde {\vet \ell}| = s}
c_{\vet \ell,\tilde {\vet \ell};\pm}^{(r-1,s)}\,
(-i \vet z)^{\vet \ell}\bar {\vet z}^{\tilde {\vet \ell}}$,
where the coefficients are complex numbers expressed as intervals,
that are explicitly computed with the help of an algebraic
manipulator\footnote{\label{nota:birkhoff_HH.mth} It is not easy to
  code with {{\it X}$\rho${\it \'o}$\nu${\it o}$\zeta$}, which is the
  algebraic manipulator we actually used for all the applications
  discussed in the present paper, because its syntax looks quite
  difficult for new users. Therefore, we think that an implementation
  of the algorithm constructing the Birkhoff normal form in the
  framework provided by {\tt Mathematica} can be more helpful for a
  reader that is interested in reproducing our results. We have
  written such a code (named {\tt birkhoff\_HH.mth}) in a version
  computing exactly the coefficients as algebraic numbers. It makes
  part of the package freely available at the web address {\tt
    http://www.mat.uniroma2.it/$\sim$locatell/CAPs/BirkCAP\_AppsA-B.zip}
  and its output is easily converted in rigorous upper bounds on the
  norms by another program (named {\tt estimates.c}). Also this
  further code is included in that same software package and is in
  charge to compute rigorously all the upper and lower bounds that are
  needed, in order to conclude the computer-assisted proof of the same
  result discussed in the present appendix.}. Moreover, the values
$\Zscr_{s}$ and $\Fscr_{s}^{(r)}$ are the rigorous upper bounds for
the norms of the corresponding Hamiltonian polynomials whose
expansions are not explicitly computed, while $\Escr$ and $a_r$ are
the parameters needed to estimate the infinite tail of terms of type
$f_{s}^{(r-1)}$ with $s>R_{\rm II}\,$. Let us recall that the values
appearing in the second and third rows of definition~\eqref{frm:array}
are expressed in logarithmic scale because some of them can eventually
exceed the ``safe range'' set $\Sgot$ of representable numbers on a
computer, as it has been widely discussed in
appendix~\ref{app:interval}.

In our opinion, the main concept to be kept in mind can be summarized
as follows. The prescriptions included in
sections~\ref{sec:birk}--\ref{sec:stime} allow to perform the $r$--th
normalization step, in such a way to explicitly determine all the
elements appearing in $\Sscr^{(r)}$, which represents $\Hscr^{(r)}$
and is a set having {\it finite} cardinality with the same structure
as that described in~\eqref{frm:array} (where $r-1$ has to be replaced
with $r$). As a short reformulation, this means that the algorithm
iteratively maps $\Sscr^{(r-1)}$ into $\Sscr^{(r)}$ and now we are
going to do it from $\Sscr^{(0)}$ to $\Sscr^{(5)}$.

\paragraph{Input.}
We focus on the Hamiltonian of the H\'enon--Heiles
model~\eqref{frm:ham-HH} with\footnote{We have made such a choice for
  the values of the angular velocities $\omega_1$ and $\omega_2\,$,
  because all the coefficients of the expansions defined by the
  normalization algorithm can be represented in the form
  $(l+m\sqrt{2})/n$, being $l,\,m,\,n\,\in\interi$. This makes easier
  the exchange of information between the codes {\tt birkhoff\_HH.mth}
  and {\tt estimates.c} as it has been described in the previous
  footnote${{\ref{nota:birkhoff_HH.mth}}\atop{\phantom{1}}}$.}
$\omega_1=1$ and $\omega_2 =-\sqrt{2}/2$.  After having rewritten it
in complex canonical variables $(-i \vet z, \bar{\vet z})$
(see~\eqref{frm:variabili-complesse}
and~\eqref{frm:da_varcanpol_a_azang} for the definition), the set
$\Sscr^{(0)}$ representing the initial Hamiltonian $\Hscr^{(0)}$ can
be written as follows:
\begin{equation}
  \label{frm:Sscr0}
\Sscr^{(0)} =
\left\{ Z_0,\  f_1^{(0)}, \ 0,\ -10^4,\ -10^4,\ -10^4, \ 0,\ 0.4424676 \right\}\ ,
\end{equation}
where
$$
Z_0 = i\, 1.00000000000000000_{\pm950} \,(-i z_1) \bar z_1
-i\,0.707106781186547573_{\pm6884} \,(-i z_2)\bar  z_2
$$ 
and 
$$
\vcenter{\openup1\jot\halign{
 \hbox {\hfil $\displaystyle {#}$}
&\hbox {\hfil $\displaystyle {#}$\hfil}
&\hbox {$\displaystyle {#}$\hfil}\cr
f_1^{(0)} &= -i \, 0.353553390593273731_{\pm3026}\, (-i z_1)^2 (-i z_2) -0.117851130197757920_{\pm 951}\ \bar z_2^3\cr
&-0.353553390593273731_{\pm 2859}\, (-i z_1)^2 \bar z_2-0.707106781186547462_{\pm 6051}\, (-i z_1) (-i z_2)\bar z_1\cr
& + i\, 0.707106781186547462_{\pm 5718} \, (-i z_1)\bar z_1  \bar z_2+i\,0.117851130197757920_{\pm 1034}\, (- i z_2)^3 \cr
&+ 0.353553390593273731_{\pm 3109}\, (-i z_2)^2 \bar z_2+ i\,0.353553390593273731_{\pm 2859}\,(- i z_2) \bar z_1^2 \cr
&-i\,0.353553390593273731_{\pm 3054}\, (- i z_2)\bar z_2^2+ 0.353553390593273731_{\pm 2693}\, \bar z_1^2 \bar z_2\ .
\cr
}}
$$
In both the expansions above we adopted the notation $c_{\pm} =
a_{\pm \sigma_a} + i\,b_{\pm\sigma_b}$ for each coefficient, where $a$
and $b$ are the central values referring to the intervals of the real
and imaginary part, while $\sigma_a$ and $\sigma_b$ allow to
determine the half-widths $\sigma_a\,\times 10^{-{\rm e}_a}$ and
$\sigma_b\,\times 10^{-{\rm e}_b}$ of those same intervals, being
${\rm e}_a$ and ${\rm e}_b$ the number of digits appearing after the
floating point in the writing of $a$ and $b$, respectively. For
instance, the last summand in the expansion of $Z_0$ reads as
$$
i\, [-0.707106781186554457\,,\,-0.707106781186540689] \,(-i z_2)\bar  z_2\ .
$$
We remark that $f_s^{(0)}=0$ $\forall\ s\ge 2$, because the
H\'enon--Heiles model is defined by a cubic Hamiltonian. This explains
why the third element of $\Sscr^{(0)}$ is equal to zero. Since we have
decided to describe the upper bounds of the norms in a logarithmic
scale (in order to maintain the agreement with more challenging
applications involving huge expansions), therefore, we have put
$\log(\Fscr_{s}^{(0)})=-10^4$ for $3\le s \le 5$ in
formula~\eqref{frm:Sscr0}; in practice, this means that we are
overestimating~$0$ with the very low positive number $e^{-1000}\simeq
10^{-434.3}$. For what concerns the last two parameters $\Escr$ and
$\log a_0$ appearing at the end of the representation
$\Sscr^{(r-1)}$~\eqref{frm:array} in the case with $r=1$, usually they
are determined taking into account both the sup norm and the
geometrical decay of the power series in a domain where the
Hamiltonian $\Hscr^{(0)}$ is an analytic function.  In this particular
case, any choice of a pair of positive values for $\Escr$ and $\log
a_0$ would be acceptable. However, since the value of the common ratio
$a_{r-1}$ (that is related to the majorants of the infinite tail of
terms appearing in the remainder of $H^{(r-1)}$) affects the next one,
i.e., $a_{r}\,$, through the relation~\eqref{frm:a_r}, we have found
reasonable to set $\Escr$ and $a_0$ in such a way to describe the
relative increase of $f_{1}^{(0)}$ with respect to $Z_{0}\,$. This is
the reason why the last two elements making part of $\Sscr^{(0)}$
in~\eqref{frm:Sscr0} have been fixed so that $\log\Escr=0$ and $\log
a_0 = \log \Fscr_{1}^{(0)} \oslash_{+} 3$, where $\log
\Fscr_{1}^{(0)}=\log_{+}(\| f_{1}^{(0)}\|)$ and the definitions of the
elementary arithmetic operators allowing to implement validated
numerics (like, e.g., $\oslash_{+}$) are given in the previous
appendix~\ref{app:interval}.

\paragraph{First normalization step: $r=1$.}
We have to put in Birkhoff normal form the cubic term. In order to do
that, we solve the homological equation $\Lie {\chi_1} Z_0 + f_1^{(0)}
= Z_1$ and, in view of equation~\eqref{frm:chi-r}, we obtain
$$
\vcenter{\openup1\jot\halign{
 \hbox {\hfil $\displaystyle {#}$}
&\hbox {\hfil $\displaystyle {#}$\hfil}
&\hbox {$\displaystyle {#}$\hfil}\cr
\chi_1&= 0.273459080339013560_{\pm 20345}\ (-i z_1)^2 (-i z_2)-i\,0.130601937481870711_{\pm 5246}\,(- i z_1)^2\bar z_2\cr
& +i\,0.999999999999999889_{\pm 40524}\,  (- i z_1) (- i z_2)\bar z_1 -i\,0.0555555555555555525_{\pm 22552}\ \bar z_2^3\cr
&-0.999999999999999889_{\pm 40080}\, (- i z_1)\bar z_1\bar z_2+0.0555555555555555525_{\pm 22934}\, (-i z_2)^3 \cr
& -i\,0.499999999999999889_{\pm 20401}\, (-i z_2)^2 \bar z_2+ 0.130601937481870711_{\pm 5246}\, (-i z_2)\bar z_1^2 \cr
& +0.499999999999999889_{\pm 20318}\, (-i z_2)\bar z_2^2-i\,0.273459080339013560_{\pm 20096}\,\bar z_1^2 \bar z_2 .
\cr
}}
$$
The new normal form term $Z_1$ is equal to $0$, due to the fact that
the normalization step is odd (see~\eqref{frm:Z_r}).  We have now to
apply the Lie series of generating function $\chi_1$ to $H^{(0)}$;
since we are considering terms up to degree $4$, the only
contributions we have to calculate are $\frac 1 2\Lie{\chi_1}^2 Z_0$
and $\Lie{\chi_1} f_1^{(0)}$; in fact, according to
formula~\eqref{frm:nuovi-pert}, they will produce terms of degree $4$,
which are collected together in the definition of
$f_2^{(1)}=\frac{1}{2}\Lie{\chi_1}^2Z_0+\Lie{\chi_1}f_1^{(0)}$.  In
order to properly describe all the Hamiltonian terms at the end of the
first normalization step, we have to update the parameters appearing
in the second and in the third row of the generic representation
described in~\eqref{frm:array}. Since we know the expansion of the
generating function $\chi_1$, we can use the formul{\ae} in
Proposition~\ref{pro:itera-stime-singole} to rigorously compute the
upper bounds $\Dscr_1$ and $\log(\Fscr_{s}^{(1)})$ for $3 \le s \le
5$; moreover, we can compute the norm of\footnote{For brevity reasons,
  here we have decided to not include the expansion of
  $f_2^{(1)}$. However, its normal form part, i.e., $Z_2$, is written
  within the following description of the second normalization step.}
$f_2^{(1)}$, that is $ \log_+(\| f_{2}^{(1)}\|)=2.446291$. We define
$\log a_1=2.599403$ by using formula~\eqref{frm:a_r}.  Therefore, the
representation of the new Hamiltonian $\Hscr^{(1)}$ can be summarized
by the following set:
$$
\Sscr^{(1)} = \left\{Z_0,  \ 0, \ f_2^{(1)},\ 6.685803, \ 8.979949, \ 11.16873,
\ 0, \ 2.599403 \right\}\ .
$$
Finally, we use formula~\eqref{frm:normasup-resto} to determine the
rigorous estimate of the sup norm of the remainder after having
completed the first normalization step:
$\log(|\Rscr^{(1)}|_\rho)=-34.71383$.

\paragraph{Second normalization step: $r =2$.}
For what concerns the second step, we proceed in an analogous way with
respect to the first one.  The homological equation we have to solve
is $\Lie{\chi_2}Z_0 + f_2^{(1)}= Z_2$. We compute the generating
function $\chi_2$ by using formula~\eqref{frm:chi-r} and the new
normal form term $Z_2$ as defined in~\eqref{frm:Z_r}. All the other
terms generated by the Lie series are of polynomial degree greater
than $4$, therefore, their Taylor expansion is not explicitly
computed. By repeating the same computations we have done for the
first step, we can write
$$
\Sscr^{(2)} = \left\{Z_0,  \ 0, \ Z_2,\ 6.685803, \ 9.014286, \ 11.55494,\ 0,\ 2.664144\right\}\ ,
$$
where
$$
\vcenter{\openup1\jot\halign{
 \hbox {\hfil $\displaystyle {#}$}
&\hbox {\hfil $\displaystyle {#}$\hfil}
&\hbox {$\displaystyle {#}$\hfil}\cr
Z_2 &=  -0.656599153958936865_{\pm 325684}\ (-i z_1)^2\bar z_1^2-0.589255650988789514_{\pm 313083}\ (-i z_2)^2 \bar z_2^2\cr
&+ 1.98564213380166632_{\pm  125900 }\ (-i z_1) (-i z_2) \bar z_1 \bar z_2\ .
\cr
}}
$$
As a new estimate for the remainder, we obtain
$\log(|\Rscr^{(2)}|_\rho)=-39.36487$. Since it is smaller with respect to
the upper bound for the same quantity at the first normalization step,
then it is convenient to further iterate the algorithm.

\paragraph{Third normalization step: $r=3$.}
As a major difference with respect to the second step, now an explicit
expansion for $f_3^{(2)}$ is not available and, therefore, the same
holds for the generating function $\chi_3\,$. Instead of computing
$\Dscr_3$ by using the expansion of $\chi_3\,$, we have to set
$\Dscr_3 = 5\Gscr_3$ (see formula~\eqref{frm:g-r} and the definition
of $\Dscr_r$ in Proposition~\ref{pro:itera-stime-singole}). Indeed,
$\Gscr_3$ can be easily computed because $\Fscr_{3}^{(2)}$ is known
and $\alpha_r$ can be determined, because it is the smallest divisor
which could appear at the third step (see~\eqref{frm:alfa-r}).
After having determined the upper bounds $\Zscr_3$ and
$\log(\Fscr_{s}^{(3)})$ for $s=4,\,5$ (by following the prescriptions
in Proposition~\ref{pro:itera-stime-singole}), we can conclude that at
the end of this normalization step the Hamiltonian is represented by
$$
\Sscr^{(3)} = \left\{Z_0,  \ 0, \ Z_2,\ 6.685803, \ 9.014286, \ 13.18247, \ 0, \ 3.937671\right\} \ .
$$
The new estimate of the remainder is
$\log(|\Rscr^{(3)}|_\rho)=-42.15919$; once again, it has decreased.

\paragraph{Fourth normalization step: $r=4$.}
We can proceed exactly in the same way as for the third step. The
representative set corresponding the new Hamiltonian $\Hscr^{(4)}$
is
$$
\Sscr^{(4)} =
\left\{Z_0,  \ 0, \ Z_2,\ 6.685803, \ 9.014286, \ 13.18247 , \ 0 , \ 4.002009 \right\} \ .
$$
The new remainder can be estimated as
$\log(|\Rscr^{(4)}|_\rho)=-41.66110$. Since it has increased, the
third normalization step can be considered as the optimal one with
respect to these parameters and the algorithm stops here.

\paragraph{End of the algorithm.}
We reconsider the set $\Sscr^{(3)}$ because it represents the
Hamiltonian at the third normalization step, which is related to the
smallest remainder. The value of $\rho_0$ that optimizes the escaping
time, as defined at the end of subsection~\ref{subsec:time-nonres}, is
$\rho_0=0.00008165$. We can finally compute the lower bound for the
stability time $T$ by using formula~\eqref{frm:time} and we obtain
$\log T=24.92920$.

\paragraph{Final comments.} It is easy to realize that the explicit
computation of the expansions for the terms belonging to the
Hamiltonians up to the degree $R_{\rm I}+2$ does not depend on the
evaluation of the upper bounds for higher order polynomials.  This
remark explains the reason why it is convenient to separate the
computer-assisted proof in two different programs, that are designed
in order to handle with the expansions and the estimates,
respectively. These two codes are included in
{\tt BirkCAP\_AppsA-B.zip} and they are in a version that is adapted
to the example discussed in the present appendix.
Such a file is conceived as a sort of supplementary
material\footnote{It is freely available at the web address already
  reported in the previous
  footnotes${{\ref{nota:lib_val_num.c},\ref{nota:birkhoff_HH.mth}}\atop{\phantom{1}}}$.}
with respect to this work, thus, it can be interesting for readers
willing to reproduce the computational algorithm in their own codes
(or to adapt our ones). For such a purpose, in
{\tt BirkCAP\_AppsA-B.zip} we have included a program computing exactly
the polynomial expansions in the framework of {\tt Mathematica}, in order
to provide a code that is rather easy to read. On the other hand,
the explicit expansions reported in this appendix have been
produced by using {{\it X}$\rho${\it \'o}$\nu${\it o}$\zeta$} for
the algebraic manipulations of polynomials with coefficients that
are complex numbers expressed as intervals. We think that this can be
useful for the initial comparisons with a new code, eventually designed
by a reader interested in much more challenging applications with
respect to the very simple example discussed in the present appendix.

\end{document}

%% file: tab.tex
\begin{table}[t] %\footnotesize
\begin{center}
\begin{tabular}{c|c|c|c|c|c|c}
 $\rho_0$ & $\rho$ & $r_{\rm opt}$ & $a_r$ & $ \log_{10}{| \Rscr^{(r_{\rm opt})}|_\rho}$ & $\log_{10}|\dot I_j|_\rho$ & $\log_{10}T$
\\
\hline
9.96e-04 & 1.00e-03 & 232 & 1.00e+03 & -1.82e+02 & -1.80e+02 & 1.72e+02 \\
1.24e-03 & 1.25e-03 & 230 & 8.02e+02 & -1.59e+02 & -1.57e+02 & 1.49e+02 \\
1.55e-03 & 1.56e-03 & 164 & 6.40e+02 & -1.42e+02 & -1.39e+02 & 1.32e+02 \\
1.94e-03 & 1.95e-03 & 144 & 5.13e+02 & -1.28e+02 & -1.26e+02 & 1.18e+02 \\
2.42e-03 & 2.44e-03 & 110 & 4.10e+02 & -1.16e+02 & -1.14e+02 & 1.07e+02 \\
3.02e-03 & 3.05e-03 & 102 & 3.28e+02 & -1.06e+02 & -1.04e+02 & 9.73e+01 \\
3.78e-03 & 3.81e-03 & 100 & 2.63e+02 & -9.63e+01 & -9.43e+01 & 8.77e+01 \\
4.72e-03 & 4.77e-03 & 100 & 2.11e+02 & -8.63e+01 & -8.43e+01 & 7.79e+01 \\
5.90e-03 & 5.96e-03 & 100 & 1.69e+02 & -7.63e+01 & -7.43e+01 & 6.82e+01 \\
7.38e-03 & 7.45e-03 & 100 & 1.35e+02 & -6.63e+01 & -6.43e+01 & 5.84e+01 \\
9.22e-03 & 9.31e-03 & 100 & 1.08e+02 & -5.64e+01 & -5.43e+01 & 4.86e+01 \\
1.15e-02 & 1.16e-02 & 74 & 8.63e+01 & -4.78e+01 & -4.59e+01 & 4.05e+01 \\
1.43e-02 & 1.46e-02 & 58 & 7.07e+01 & -4.18e+01 & -4.00e+01 & 3.48e+01 \\
1.79e-02 & 1.82e-02 & 52 & 5.66e+01 & -3.67e+01 & -3.49e+01 & 3.00e+01 \\
2.23e-02 & 2.27e-02 & 52 & 4.49e+01 & -3.13e+01 & -2.96e+01 & 2.49e+01 \\
2.79e-02 & 2.84e-02 & 48 & 3.57e+01 & -2.67e+01 & -2.50e+01 & 2.05e+01 \\
3.46e-02 & 3.55e-02 & 38 & 2.84e+01 & -2.27e+01 & -2.11e+01 & 1.68e+01 \\
4.30e-02 & 4.44e-02 & 30 & 2.32e+01 & -1.97e+01 & -1.82e+01 & 1.43e+01 \\
5.36e-02 & 5.55e-02 & 26 & 1.86e+01 & -1.71e+01 & -1.56e+01 & 1.19e+01 \\
6.70e-02 & 6.94e-02 & 26 & 1.49e+01 & -1.42e+01 & -1.28e+01 & 9.30e+00 \\
8.37e-02 & 8.67e-02 & 26 & 1.15e+01 & -1.14e+01 & -9.94e+00 & 6.65e+00 \\
\end{tabular}
\end{center}
\caption[]{In this table we report the results obtained for the H\'enon-Heiles model with frequencies $\omega_1 = 1$ and $\omega_2 = -(\sqrt 5 -1)/2$.}
\label{tab:chronos+stime}
\end{table}

%% file: Jupcomp.tex
\begin{table}
\begin{minipage}{0.5\linewidth}
\begin{center}
\begin{tabular}{c|c|c}
$\rho_0^2$ & $\rho^2$ & $T$ \\
\hline
2.49e-04 & 2.59e-04 & 6.36e+08 \\
2.47e-04 & 2.57e-04 & 1.01e+09 \\
\end{tabular}
\end{center}
\end{minipage}
\hfill
\begin{minipage}{0.5\linewidth}
\begin{center}
\begin{tabular}{c|c|c|c}
$\rho_0^2$ & $({\rho^*_2})^2$ & $\rho^2$ & $T$ \\
\hline
2.05e-04 & 1.83e-04 & 2.07e-04 & 5.93e+08 \\
2.02e-04 & 1.80e-04 & 2.04e-04 & 7.23e+08 \\
\end{tabular}
\end{center}
\end{minipage}
\caption[]{Comparison for the estimates on the stability time between
  the non-resonant and resonant Birkhoff normal forms. The Jupiter case
  ($\mu\simeq 0.000954$) with $T_{ \rm e. l. t.}  \simeq 5\times 10^8$.}
\label{tab:jupcomp}
\end{table}

%% file: Uracomp.tex
\begin{table}[t]
\begin{minipage}{0.45\linewidth}
\begin{center}
\begin{tabular}{c|c|c}
$\rho_0^2$ & $\rho^2$ & $T$ \\
\hline
8.30e-05 & 8.80e-05 & 6.03e+07 \\
8.13e-05 & 8.63e-05 & 1.44e+08 \\
\end{tabular}
\end{center}
\end{minipage}
\hfill
\begin{minipage}{0.55\linewidth}
\begin{center}
\begin{tabular}{c|c|c|c}
$\rho_0^2$ & $({\rho^*_2})^2$ & $\rho^2$ & $T$ \\
\hline
9.23e-04 & 7.57e-04 & 9.24e-04 & 7.18e+07 \\
9.04e-04 & 7.44e-04 & 9.05e-04 & 1.27e+08 \\
\end{tabular}
\end{center}
\end{minipage}
\caption[]{As in Table~\ref{tab:jupcomp} for the
  Uranus case ($\mu\simeq 4.36\times 10^{-5}$) with $T_{ \rm e. l. t.}
  \simeq 6\times 10^7$.}
\label{tab:Uranus}
\end{table}

%% file: Marcomp.tex
\begin{table}[t]
\begin{minipage}{0.45\linewidth}
\begin{center}
\begin{tabular}{c|c|c}
$\rho_0^2$  & $\rho^2$ & $T$ \\
\hline
7.36e-06 & 7.84e-06 & 3.09e+09 \\
7.22e-06 & 7.69e-06 & 6.15e+09 \\
\end{tabular}
\end{center}
\end{minipage}
\hfill
\begin{minipage}{0.55\linewidth}
\begin{center}
\begin{tabular}{c|c|c|c}
$\rho_0^2$  & $({\rho^*_2})^2$ & $\rho^2$ & $T$ \\
\hline
1.28e-04 & 1.08e-04 & 1.28e-04 & 3.87e+09 \\
1.27e-04 & 1.07e-04 & 1.27e-04 & 5.86e+09 \\
\end{tabular}
\end{center}
\end{minipage}
\caption[]{As in Table~\ref{tab:jupcomp} for the Mars case ($\mu\simeq
  3.21\times 10^{-7}$) with $T_{ \rm e. l. t.}  \simeq 3 \times
  10^9$.}
\label{tab:Mars}
\end{table}

%% file: Jancomp.tex
\begin{table}[t]
\begin{minipage}{0.45\linewidth}
\begin{center}
\begin{tabular}{c|c|c}
$\rho_0^2$ & $\rho^2$ & $T$ \\
\hline
6.00e-07 & 6.37e-07 & 3.10e+12 \\
5.89e-07 & 6.24e-07 & 5.40e+12 \\
\end{tabular}
\end{center}
\end{minipage}
\hfill
\begin{minipage}{0.55\linewidth}
\begin{center}
\begin{tabular}{c|c|c|c}
$\rho_0^2$ & $({\rho^*_2})^2$ & $\rho^2$ & $T$ \\
\hline
1.18e-05 &1.10e-05 & 1.18e-05 & 3.50e+12 \\
1.15e-05 & 1.08e-05 & 1.15e-05 & 6.83e+12 \\
\end{tabular}
\end{center}
\end{minipage}
\caption[]{As in Table~\ref{tab:jupcomp} for the Janus case
  ($\mu\simeq 3.36\times 10^{-9}$) with $T_{ \rm e. l. t.}  \simeq 3
  \times 10^{12}$.}
\label{tab:Janus}
\end{table}